\documentclass[aps, pre, twocolumn, superscriptaddress, reprint]{revtex4-1}

\usepackage{amsfonts, amsmath, bm, graphicx}
\usepackage{soul}
\usepackage{color}
\def\vec#1{{\bm{#1}}}
\def\mat#1{{\hat{\vec{#1}}}}
\def\H{{\mathcal{H}}}
\def\t#1{{\mathrm{#1}}}
\DeclareUnicodeCharacter{2212}{-}

\usepackage{siunitx}
\begin{document}

\title{Theory of three-magnon interaction in a vortex-state magnetic nanodot}

\author{Roman Verba} \email[corresponding author, e-mail: ]{verrv@ukr.net}
\affiliation{Institute of Magnetism, Kyiv 03142, Ukraine}

\author{Lukas K\"{o}rber}
\affiliation{Helmholtz-Zentrum Dresden-Rossendorf, Institute of Ion Beam Physics and Materials Research, 01328 Dresden, Germany}
\affiliation{Fakult\"at Physik, Technische Universit\"{a}t Dresden, 01062 Dresden, Germany}

\author{Katrin Schultheiss}
\affiliation{Helmholtz-Zentrum Dresden-Rossendorf, Institute of Ion Beam Physics and Materials Research, 01328 Dresden, Germany}

\author{Helmut Schultheiss}
\affiliation{Helmholtz-Zentrum Dresden-Rossendorf, Institute of Ion Beam Physics and Materials Research, 01328 Dresden, Germany}
\affiliation{Fakult\"at Physik, Technische Universit\"{a}t Dresden, 01062 Dresden, Germany}

\author{Vasil Tiberkevich}
\affiliation{Department of Physics, Oakland University, Rochester, Michigan 48309, USA}

\author{Andrei Slavin}
\affiliation{Department of Physics, Oakland University, Rochester, Michigan 48309, USA}

\date\today


\begin{abstract}
We use vector Hamiltonian formalism (VHF) to study theoretically three-magnon parametric interaction (or three-wave splitting) in a magnetic disk existing in a magnetic vortex ground state. The three-wave splitting in a disk is found to obey two selection rules: (i)  conservation of the total azimuthal number of the interacting spin-wave modes, and (ii) inequality of the radial numbers of the resultant modes, if the directly excited original mode is radially symmetric (i.e. if the azimuthal number of the directly excited mode is $m = 0$). The selection rule (ii), however, is relaxed in sufficiently small magnetic disks, due to the influence of the vortex core. We also found, that the efficiency of the three-wave splitting of the directly excited mode strongly depends on the azimuthal and radial mode numbers of the resultant modes. This property becomes qualitatively important in the case when several different splitting channels (several pairs of resultant modes) approximately satisfy the resonance condition for the splitting. The good agreement of the VHF analytic calculations with the experiment and micromagnetic simulations proves the capability of the VHF formalism to predict the  actual experimentally realized splitting channels, and the magnitude of the driving field thresholds for the three-wave splitting.
\end{abstract}

\maketitle


\section{Introduction}

The intrinsic nonlinearity of magnetization dynamics in ferromagnetic materials leads to a wide variety of nonlinear phenomena, which can be observed in experiment and utilized in practice \cite{Kosevich_Book, Gurevich_Book1996, Bertotti_Book}. At relatively low driving field powers the observed nonlinear magnetization dynamics can be, often, interpreted as an interaction of multiple linear spin-wave (SW) eigenmodes (or magnons) -- i.e. as multi-magnon interaction processes \cite{Schlomann_PR1959, Lvov_Book1994, Safonov_Book}. The most important among these interaction processes are the lowest-order three-magnon and four-magnon interactions, even though there are cases when higher-order processes can become important as well \cite{Gottlieb_JAP1962}. Three-magnon processes cause the, so-called, first-order Suhl instability of uniform magnetization precession \cite{White_PR1963, Patton_PSS1979, Nazarov_JMMM2002}, and nonlinear decay of propagating SWs \cite{Boardman_PRB1988, Liu_PRB2019}. Four-magnon processes, some of which are always allowed, are responsible for the nonlinear shift of the SW frequency, the foldover effect \cite{Suhl_JAP1960, Gottlieb_JAP1962, Gui_PRB2009}, phase mechanism of the parametric resonance saturation \cite{Zakharov_UFN1975, Lvov_Book1994}, and the formation of SW envelope solitons \cite{Kalinikos_JETP1988, Chen_PRL1993, Melkov_JAP2001}.

Nonlinear SW interaction has been studied for a long time in bulk samples and in thin ferromagnetic films \cite{White_PR1963, Patton_PSS1979, Nazarov_JMMM2002, Boardman_PRB1988, Livesey_PRB2007}. However, in magnetic nanostructures the properties of the multi-magnon interaction could differ substantially from the properties of similar processes  in the bulk magnetic samples. First, the quantization of the frequencies and wavevectors of the SW eigenmodes due to the spatial confinement in nanostructures makes the exact fulfillment of the resonance conditions for a particular magnon interaction process difficult to achieve. Thus, instead of resonant nonlinear processes, common for bulk magnetic materials, the nonresonant processes are often realized in finite-size magnetic nanostructures \cite{Melkov_MagLet2013, Slobodianiuk_MagLett2019}. In particular, the discreteness of the SW spectrum manifests itself in the appearance of specific features of the nonlinear ferromagnetic resonance  \cite{Melkov_MagLet2013, Slobodianiuk_MagLett2019}, in the strong frequency-dependent nonlinear enhancement of the SW damping \cite{Barsukov_SciAdv2019}, in the possibility of the excitation of stable large-angle magnetization precession \cite{Kobljanskyj_SR2012},  etc.. Second, a spatial nonuniformity of the magnetization ground state (e.g., vortex ground state), and the corresponding specific structure of the linear SW modes result in the selection rules for the three-magnon and higher-order processes, which are specific for a magnetic nanostructure having a particular shape and a particular magnetic ground state \cite{Camley_PRB2014, Schultheiss_PRL2019}.

In our recent paper \cite{Schultheiss_PRL2019}, we  observed experimentally the three-magnon splitting of a directly excited dipolar SW mode in a vortex-state magnetic disk. The application of a sufficiently large microwave magnetic field with an out-of-plane polarization leads to the splitting of a directly exited radial SW mode into a pair of azimuthal SW modes. This experiment allowed us, for the first time,  to observe the dynamic SW modes of a magnetic vortex with unusually high azimuthal numbers. These magnon modes resemble the ``whispering gallery modes''\cite{Reyleigh_1914} previously studied in other physical systems,  and may be interesting for applications with whispering gallery modes of other nature (e.g., photonic) in various hybrid structures.

\begin{figure}
 \includegraphics[width=\columnwidth]{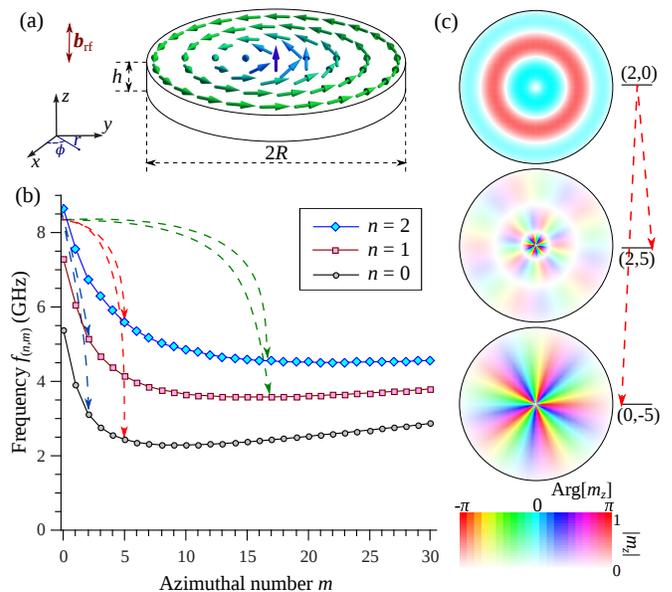}
 \caption{(a) Sketch of a circular magnetic dot in a vortex ground state; (b) Calculated SW spectrum of a  permalloy dot having diameter $2R = 5.1\,\t{\mu m}$ and thickness $t = 50\,\t{nm}$; solid lines are guides to the eye; dashed lines show possible splitting channels, which are close to the three-magnon splitting resonance condition at the excitation frequency of 8.3~GHz; (c) Spatial profiles of directly excited and split SW modes  at the excitation frequency of 8.3~GHz.  Profiles are obtained using numerical calculations (see text), where the magnitude and relative phase of the magnetization oscillations are coded using intensity and color scale, respectively.}\label{f:1}
\end{figure}

In our current work, we study theoretically the three-magnon splitting process in a magnetic nanodot existing in a vortex ground state. The main goal of this study is to formulate the selection rules for three-magnon splitting, i.e., to find out which splitting processes are allowed and which ones are not. Our second goal is to derive expressions for the coefficients of the three-magnon interaction (often referred to as the ``three-magnon coefficients''). A quantitative  knowledge of these coefficients is important not only for the calculation of the power thresholds of the three-magnon splitting processes, but also for the determination of the actual splitting channels which will be observed in an experiment. Indeed, the SW spectrum of a vortex-state disk with a micrometer-sized diameter is quite dense, as is clear from the example shown in Fig.~\ref{f:1}(b). Consequently, the resonance condition for the splitting (energy conservation rule $\omega_0 = \omega_1 + \omega_2$) could be approximately (to the accuracy of the frequency linewidth of the initially excited SW mode) satisfied for several pairs of the resultant (split) SW modes simultaneously (see arrows in Fig.~\ref{f:1}(b)). In such a case, the actual splitting channel is chosen as a channel having the largest three-magnon coefficient among all the possible channels which approximately satisfy the three-magnon resonance condition. Additionally, a quantitative knowledge of the three-magnon coefficients could be important when designing experiments on stimulated splitting, switching between the splitting channels, etc. .

The paper is organized as follows. In Sec.~\ref{s:VHF}, we briefly describe the basics of the vector Hamiltonian formalism for nonlinear SW interaction \cite{Tyberkevych_ArXiv, Dzyapko_PRB2017}. The selection rules and the general expression for the three-magnon coefficients are derived in Sec.~\ref{s:theory}. Results of the  numerical simulations of three-magnon splitting process and a comparison of the VHF analytical results to the experimental and micromagnetic simulations results are presented in Sec.~\ref{s:numerical}. Finally, in Sec.~\ref{s:summary} we present a summary of our work.

\section{Basic equations of vectorial Hamiltonian formalism}\label{s:VHF}

Nonlinear interaction between SW modes is commonly studied theoretically within the Hamiltonian approach for magnetization dynamics. The main point of the Hamiltonian approach is the representation of the components of dynamical magnetization in the form of canonical variables, and the consequent transformation of the total magnetic energy of a system  into a Hamiltonian function expanded on a power series of canonical variables. In almost all the previous papers based on the Hamiltonian approach to magnetization dynamics the authors used the scalar canonical variables $a$, $a^*$, which are related to the components of the magnetization vector by a classical analog of the first Holstein-Primakoff transformation \cite{Schlomann_PR1959}. This ``scalar Hamiltonian approach'' was successfully used for the investigation of nonlinear SW interactions in bulk magnetic samples, thin magnetic films (see, e.g., \cite{Lvov_Book1994} and references therein) and even some examples  of magnetic nanostructures \cite{Camley_PRB2014, Verba_SciRep2016}.  Using the approach of an ``effective SW tensor'' \cite{Nazarov_JMMM2002}, it became possible to derive rather general expressions for the nonlinear SW coefficients describing a wide variety of spin wave self-interactions \cite{Krivosik_PRB2010, Verba_PRB2019}, which greatly simplify calculation of nonlinear coefficients. However, formalism from Refs.~\onlinecite{Krivosik_PRB2010, Verba_PRB2019} could be straightforwardly applied only to the  magnetization dynamics in the ferromagnetic samples existing in the \emph{saturated (quasi-uniform) }ground state and having spin wave eigenmodes similar to plane waves.

In modern nano-magnetism one often has to deal with magnetic nanostructures, that exist in a spatially nonuniform (e.g. vortex)  ground  state, and/or have non-plane-wave-like SW eigenmodes. In such a case scalar Hamiltonian formalism encounters serious difficulties. First, is the need to introduce a local coordinate system having $z$-axis in the direction of static magnetization \cite{Abyzov_JETP1979}, which could be not an easy task for a complex static magnetization profile. Second is that the standard scheme of the scalar Hamiltonian formalism -- conversion the magnetization to canonical variables $a(\vec r, t)$ and $a^*(\vec r, t)$, expansion of the canonical variables into a series of complete basis function (usually, plane waves, but could be any other basis) and subsequent diagonalization of quadratic part of the Hamiltonian using $u$-$v$ Bogoliubov transformations -- relies on the assumption that the ellipticity  of the excited SW modes is spatially uniform, i.e., that the profile of an SW mode can be expressed as $\vec m(\vec r) = \vec m f(\vec r)$.  This is, however, is not always the case even in simple geometries \cite{Wang_PRL2019}. An attempt to account for a different spatial distribution of dynamic magnetization components into Hamiltonian was done in \cite{Galkin_PRB2006}, although this has never been done for nonlinear interactions in magnetization dynamics. These points, together with the need to start consideration from the very beginning for each new static magnetization distribution or set of SW eigenmodes make application of a scalar Hamiltonian formalism to nonlinear magnetization dynamics in magnetic nanostructures very cumbersome, despite of no fundamental limitations.

The viable alternative is to use a recently developed ``vectorial Hamiltonian formalism'' \cite{Tyberkevych_ArXiv}, which can easily deal with the spatial nonuniformity of both static and dynamic magnetization.  We will use this vectorial Hamiltonian formalism in our current work. The main novel feature of the vectorial Hamiltonian approach is the mapping of a dynamics of a constant-amplitude three-dimensional magnetization vector on a unit sphere $|\vec M(\vec r, t)|/M_s = 1$ to the dynamics of a two-dimensional vector of dynamic magnetization on a plane disk.  This mapping is analogous to the Lambert azimuthal equal-area projection \cite{Snyder_1987}, and is given by the following vectorial equation:
  \begin{equation}\label{eq:trafo}
 \frac{\vec M(\vec r, t)}{M_s} = \left(1-\frac{|\vec s(\vec r, t)|^2}{2} \right) \vec \mu(\vec r) + \sqrt{1-\frac{|\vec s(\vec r, t)|^2}{4}} \vec s (\vec r,t) \ .
  \end{equation}
Here $\vec\mu (\vec r) = \vec M_0(\vec r)/M_s$ is the spatial distribution of the normalized static magnetization, $M_s$ is the saturation magnetization and $\vec s(\vec r, t)$  is the normalized dynamic magnetization, which is perpendicular to the static one, $\vec s \bot \vec \mu$. The dynamic magnetization can be expanded in a series of linear SW eigenmodes $\vec s_\nu$ of the system:
  \begin{equation}\label{e:s-exp}
   \vec s(\vec r, t) = \sum\limits_\nu \left(c_\nu(t) \vec s_\nu (\vec r) + \t{c.c.} \right) \,,
  \end{equation}
where $c_\nu$ are the complex amplitudes of the SW eigenmodes. The spatial profiles  $\vec s_\nu$ and the frequencies $\omega_\nu$ of the SW linear eigenmodes modes are the solution of the linearized Landau-Lifshits equation \cite{Naletov_PRB2011}:
  \begin{equation}\label{e:eigeneq}
   -i\omega_\nu \vec s_\nu = \vec \mu \times \mat\Omega \cdot \vec s_\nu \,,
  \end{equation}
with the operator $ \mat \Omega$ given by:
  \begin{equation}
   \mat \Omega = \gamma B \mat I + \omega_M \mat N \,,
  \end{equation}
where $B$ is the projection of the static internal magnetic field on the direction of static magnetization, $\mat I$ is the unit matrix, $\omega_M = \gamma \mu_0 M_s$, and $\mat N$ is the tensor describing magnetic self-interactions, such as  exchange, magnetodipolar, anisotropy, etc. (explicit expressions are given below). The solution of Eq.~\eqref{e:eigeneq} gives SW the spatial profiles of SW eigenmodes to the accuracy of an arbitrary multiplier. Therefore, within the vector Hamiltonian formalism the mode profiles should be normalized as follows:
  \begin{equation}\label{e:norm}
    \frac{i}{V_d} \int \vec s_\nu^* \cdot \vec\mu \times \vec s_\nu d\vec r = 1 \,,
  \end{equation}
where the integration goes over all the sample volume $V_d$. This normalization ensures that quadratic part of the normalized magnetic energy assumes  a standard Hamiltonian form in terms of the SW mode amplitudes:  $\H^{(2)} = (1/2) \sum_\nu |c_\nu|^2\omega_\nu$ (we use here a common definition of an SW Hamiltonian $\H = \gamma E/(M_s V_d)$ which is measured in the units of frequency \cite{Krivosik_PRB2010}, where $E$ is the total magnetic energy).

The three-wave term of the SW Hamiltonian function can be expressed as:
  \begin{equation}\label{e:H3-gen}
   \H^{(3)} = -\frac{\omega_M}{2V_d} \int (|\vec s|^2 \vec \mu) \cdot \mat N \cdot \vec s d\vec r \ .
  \end{equation}

Using the eigenmode expansion \eqref{e:s-exp} we can represent \eqref{e:H3-gen} in the standard form:
  \begin{equation}
  \begin{split}
       \H^{(3)} & = \frac13 \sum\limits_{123} \left(U_{123}c_1 c_2 c_3 + \t{c.c.} \right) \\ &+ \sum\limits_{123} \left(V_{12,3}c_1 c_2 c_3^* + \t{c.c.} \right) .
  \end{split}
  \end{equation}

In our current work we are interested \textit{only} in the second term of the above equation, as this term  describes a splitting of an SW mode ``3'' into a pair of SW modes ``1'' and ``2'',  and the reverse mode confluence process, denoted by the short notation $3 \to (1+2)$. The first term describes the so-called ``explosive'' instability of SW modes (nucleation or annihilation of three SW modes in vacuum) which can never be resonant in an equilibrium magnetic medium.

The coefficient of the three-wave splitting/confluence interaction can be expressed as:
  \begin{equation}\label{e:V123-gen}
  \begin{split}
   V_{12,3} &= -\frac{\omega_M}{2V_d} \int \left((\vec s_2 \cdot \vec s_3^*) \vec\mu \cdot \mat N \cdot \vec s_1 \right. \\
   &+ \left. (\vec s_1 \cdot \vec s_3^*) \vec\mu \cdot \mat N \cdot \vec s_2 + ({\vec s_1} \cdot \vec s_2) \vec\mu \cdot \mat N \cdot \vec s_3^*\right)d\vec r \ .
  \end{split}
  \end{equation}
This last equation is  convenient to use for both analytical and numerical analysis of three-wave interaction in magnetism. It should be noted, that the SW mode profiles $\vec s_\nu$ determining through \eqref{e:V123-gen} the magnitude of the three-wave interaction coefficient could be obtained not only by analytical or numerical solution of Eq.~\eqref{e:eigeneq}, but also by other methods, e.g.  using direct micromagnetic simulations.


\section{Selection rules and three-magnon interaction efficiency}\label{s:theory}

We consider nonlinear SW interaction in a thin cylindrical magnetic dot of the thickness $h$ and radius $R$ ($h \ll R$), a sketch of which is shown in Fig.~\ref{f:1}(a). The dot exists in a vortex ground state. In the polar coordinate system $(r,\phi,z)$, distribution of static magnetization of the dot is expressed as $\vec\mu = (0, \chi \sin\theta(r),p\cos\theta(r))$, where $p$ and $\chi$ are the vortex polarity and chirality, respectively. For definiteness, below we use $\chi = p = +1$. The function $\theta(r)$ describes the profile of the vortex core \cite{Usov_JMMM1993}, and it is equal to $\theta(r) = \pi/2$ away from the vortex core and $\theta(0) = 0$ at the core center.

The SW spectrum of a vortex-state dot consists of a gyrotropic mode and a set of magnetostatic modes \cite{Guslienko_JNN2008}. These magnetostatic modes in the case of a thin dot have a form of waves traveling along the azimuthal direction (around the vortex core), and are characterized by their radial index $n = 0,1,2,...$ (number of nodes in the radial direction) and azimuthal index $m = 0, \pm 1,\pm 2,..$, which describes the phase shift accumulation during one turn around the core (in $2\pi$ units). The spatial profiles of several of the magnetostatic modes are shown in Fig.~\ref{f:1}c. Of course, there are also SW modes having  a nonuniform thickness profile and characterized by the thickness index $l>0$. In our case of a thin dot, these modes have much larger frequencies, and will not be considered. However, the  analysis of the nonlinear interaction between these higher-order thickness SW modes can be done in the same manner as for the modes that are uniform along the thickness direction.

\subsection{Case of a large dot}

First, let us consider the case of a relatively large dot with a radius much larger than the size of the vortex core. As the size of the vortex core is typically of the order of $10-20\,\t{nm}$, this approximation holds for dots with radii of several hundreds of nanometers and more. In this case, one can completely neglect the presence of the vortex core, and approximate the static magnetization distribution as  $\vec\mu = \vec e_\phi$. Also, the gyrotropic mode can be disregarded, as it is localized in the vicinity of the core, and has a frequency that is much lower than the frequencies of all the other modes. The profiles of the magnetostatic SW modes are derived as $\vec s_\nu = (s_{\nu,r} (r), 0, s_{\nu,z}(r)) e^{im_\nu \phi}$.
It is important to note, that in a large dot SW modes with opposite azimuthal indices $+m$ and $-m$ are degenerate in frequency, and have the same radial profiles, i.e., $s_{(n,m),r}(r) = s_{(n,-m),r}(r)$ and $s_{(n,m),z}(r) = s_{(n,-m),z}(r)$. The only exception are the modes with $m=\pm 1$, for which this degeneracy is lifted due to the hybridization with the gyrotropic mode \cite{Guslienko_PRL2008}, leading to a nonzero frequency splitting, and a small difference in their profiles even in a relatively large vortex-state dot.

The above described general expressions for the SW spatial profiles and the distributions of static magnetization are sufficient to analyze the three-magnon interaction. As it is clear from Eq.~\eqref{e:V123-gen}, the contributions of different magnetic interactions to the three-magnon coefficients are additive, which allows us to consider the exchange contribution $V_{12,3}^\t{(ex)}$ and the dipolar contribution $V_{12,3}^\t{(dip)}$ separately. The total three-magnon-interaction efficiency is simply the sum of these contributions: $V_{12,3} = V_{12,3}^\t{(ex)} + V_{12,3}^\t{(dip)}$.

\textbf{Exchange contribution.} The tensor operator of nonuniform exchange is given by $\mat N_\t{ex} = -\lambda^2 \mat I \nabla^2$, where $\lambda$ is the exchange length of the magnetic material \cite{Nazarov_JMMM2002, Naletov_PRB2011}. Note, that this expression should be applied to magnetization components in the \emph{Cartesian} coordinate system. Since we use \emph{polar} magnetization components, the coordinate system transformation should be applied, which yields the following operator written in polar coordinates:
  \begin{equation}\label{e:N_exch}
   \mat N_\t{ex}^{\t{pol}} = - \lambda^2 \left[\mat I \nabla^2 + \frac{1}{r^2} \left(
   \begin{array}{ccc}
    -1 & -\partial_\phi & 0 \\
    \partial_\phi & -1 & 0 \\
    0 & 0 & 0
   \end{array}
   \right)\right] \ .
  \end{equation}

Using this expression in Eq.~\eqref{e:V123-gen}, we obtain the following  exchange contribution to the three-wave interaction coefficient:
  \begin{equation}\label{e:V123-exch}
  \begin{split}
   V_{12,3}^\t{(ex)} &= \frac{i \omega_M \lambda^2}{R^2} \int\limits_0^R \frac{dr}{r} \left[m_1 s_{2,z} \left(s_{1,r} s_{3,z}^* - s_{1,z} s_{3,r}^* \right) \right. \\
   &+ \left. m_2 s_{1,z} \left(s_{2,r} s_{3,z}^* - s_{2,z} s_{3,r}^* \right) \right] \Delta(m_1+m_2-m_3) \ .
  \end{split}
  \end{equation}
Here, $\Delta$ is the Kronecker delta  which gives the first selection rule: $m_3 = m_1 + m_2$. This is, in fact, the conservation of the total azimuthal number in the three-wave splitting process which reflects the conservation of the angular momentum. In a general case, this is the only restriction imposed on the vortex-state dot dynamic modes which can be involved in the three-wave interaction (splitting).

In the case of splitting of radial modes characterized by $m_3 = 0$ (or reverse confluence process into a mode with $m_3 = 0$), which is the case realized in our experiment, the azimuthal numbers of the split modes are opposite, $m_1 = -m_2= m$, and Eq.~\eqref{e:V123-exch} is simplified to
  \begin{equation}\label{e:V123-exch-m0}
   V_{12,3}^\t{(ex)} = \frac{i m \omega_M \lambda^2}{R^2} \int\limits_0^R \frac{dr}{r} \left(s_{1,r} s_{2,z} - s_{1,z} s_{2,r} \right) s_{3,z}^* \ .
  \end{equation}
From this expression it is clear, that if the split modes have the same radial index $n$, i.e., have the same profiles $s_r(r)$ and $s_z(r)$, the efficiency of the three-wave interaction is zero, $V_{12,3} = 0$. Thus, the splitting of the $m_3 = 0$ mode obeys an additional selection rule, requiring that the radial indices of the resultant (split) modes are different, $n_1\neq n_2$. The only exception is the case when $m= \pm 1$ modes, since these modes have different spatial profiles due to their hybridization with the gyrotropic mode. However, in a sufficiently large dot this difference is small, leading to a relatively small exchange contribution to the  three-wave interaction efficiency.

It should be noted, that in a uniformly magnetized sample the exchange interaction does not contribute at all to the three-wave interaction efficiency. However, a non-uniformity of the static magnetization distribution relaxes this restriction, and the exchange contribution to the three-wave coefficients becomes non-zero. For vortex-state dots, this contribution is proportional to $(\lambda/R)^2$ , and, typically, is significantly smaller than the dipolar contribution, calculated below.

\textbf{Dipolar contribution.} The tensor operator describing the magneto-dipolar interaction is expressed via the magnetostatic Green's function $\mat G$:
  \begin{equation}
   \mat N_\t{dip} \cdot \vec s = \int \mat G (\vec r, \vec r') \cdot \vec s(\vec r') d \vec r'.
  \end{equation}
For a thin dot having a spatially uniform distribution of both the static and dynamic magnetization across the dot thickness, it follows that  $G_{rz} = G_{zr} = G_{\phi z} = G_{z\phi} = 0$ \cite{Guslienko_JAP2000}. From Eq.~\eqref{e:V123-gen}, it is clear that the only component which contributes to three-wave interaction efficiency in the approximation of a relatively large dot is the off-diagonal component $G_{\phi r}$, since $\vec\mu \cdot \mat G \cdot \vec s = G_{\phi r} s_r$. This component can be expressed as \cite{Guslienko_JAP2000}:
  \begin{equation}
   G_{\phi r} (\vec r, \vec r') = \frac{i}{2\pi r} \sum\limits_m e^{im(\phi - \phi')} \int dk \frac{f(kh)}k J'_{m}(kr') J_m(kr) \,,
  \end{equation}
where the function $f(x) = 1 - (1-e^{-|x|})/|x|$, and $J'_m(kr) = dJ_m(kr)/dr = (k/2) \left(J_{m-1}(kr) - J_{m+1}(kr) \right)$ is the derivative of a Bessel function of the first kind $J_m$. Using this expression, one can find the dipolar contribution to the three-wave interaction efficiency, which, in a general case, is equal to:
  \begin{equation}\label{e:V123-dip-gen}
    \begin{split}
     V_{12,3}^\t{(dip)} & = - \frac{i \omega_M}{R^2} \int dr \int r'dr' \int dk \frac{f(kh)}{k} \\
     & \times \left[ m_1 J'_{m_1}(kr') J_{m_1}(kr) s_{1,r}(r') \left(\vec s_2(r) \cdot \vec s_3^*(r) \right) \right.  \\
     & + m_2 J'_{m_2}(kr') J_{m_2}(kr) s_{2,r}(r') \left(\vec s_1(r) \cdot \vec s_3^*(r) \right)  \\
     & - \left. m_3 J'_{m_3}(kr') J_{m_3}(kr) s_{3,r}^*(r') \left(\vec s_1(r) \cdot \vec s_2(r) \right) \right] \\
     & \times \Delta(m_1+m_2-m_3 ) \ .
    \end{split}
  \end{equation}
Here for brevity,  we use the notation $\vec s(r) = (s_r(r), 0, s_z(r))$, which describes the radially-dependent part of the mode profile. Similar to the exchange contribution, in a general case, the only selection rule is the one imposed on the  azimuthal indices of the SW modes, and it requires conservation of the total azimuthal number.

For the splitting of the azimuthally symmetric mode ($m_3 = 0$, and, consequently, $m_1 = -m_2 = m$), Eq.~\eqref{e:V123-dip-gen} can be significantly simplified to:
  \begin{equation}\label{e:V123-dip-m0}
   \begin{split}
      V_{12,3}^\t{(dip)} &= - \frac{i m \omega_M}{2R^2} \int dr \int r'dr' \int dk f(kh) \\
      &\times \left(J_{m-1}(kr') - J_{m+1}(kr') \right) J_{m}(kr) \\
      &\times \left[s_{1,r}(r') \vec s_2(r) \cdot \vec s_3^*(r) - s_{2,r}(r') \vec s_1(r) \cdot \vec s_3^*(r) \right] \ .
    \end{split}
   \end{equation}
From the last term in the above expression it is clear, that if $\vec s_1(r) = \vec s_2(r)$, the three-wave interaction efficiency is equal to zero. As a consequence, the dipolar contribution results in the same selection rule as the exchange one -- if $m_3 = 0$, then $n_1 \neq n_2$. As before, the only exception is for the  modes with $|m| = 1$, which are not degenerate due to the hybridization with the gyrotropic mode. However, in a large dot, the difference in mode profiles caused by the hybridization and, consequently, the contribution to the three-wave coefficient, are small, so that the splitting process $(n_3,0)\to (n,1) + (n,-1)$ would be hard to observe in experiment.

In summary, we can conclude that in the case of a relatively large vortex-state dot the three-wave splitting process into frequency-degenerate modes is impossible. If a directly excited mode is not radially symmetric, $m_3 \neq 0$, than the resultant (split) modes differ by the modulus of azimuthal number, $|m_1| \neq |m_2|$, as it follows from the conservation of azimuthal number. At the same time, in this case there are no restrictions on the radial numbers of the modes involved in the splitting process. If a directly excited mode is radially symmetric, $m_3 = 0$, than the split modes should differ by the radial number, $n_1 \neq n_2$, while there are no restrictions on the relation between the radial number of directly excited  mode and the radial numbers  of the split modes.

\subsection{Effect of the vortex core}

Let us now consider how the presence of the vortex core affects the three-wave interaction efficiency. For this study, we have to use the full expressions for the spatial distribution of the static magnetization $\vec\mu = [0,\sin \theta(r), \cos \theta(r)]$, and for the SW mode spatial profiles, which can be expresses as $\vec s = [s_r(r), -s_\xi(r) \cos\theta(r), s_\xi(r) \sin\theta(r)] e^{im\phi}$. Here, $\xi$ is the local coordinate axis which is perpendicular to both $\vec\mu$ and $\vec e_r$. Using these expressions, one can calculate the three-wave interaction efficiency in the same manner as presented above.

In this general case, the expressions for the splitting efficiency too cumbersome, even for the case $m_3 = 0$, and we do not present them here. Simultaneously, we would like to point out, first, that the selection rule for the azimuthal indices $m_3 = m_1 + m_2$ is not changed by the influence of the vortex core. Indeed, this rule comes from the integration $\exp[i(m_1+m_2-m_3)\phi]$ over the azimuthal coordinate, and the vortex core does not introduce any additional dependence of the static or dynamic magnetization on the azimuthal coordinate $\phi$. Second, our calculations show that the effect of the vortex core relaxes the selection rule $n_1 \neq n_2$ for $m_3 = 0$. The corresponding contribution to  the three-wave interaction efficiency is found to be proportional to $\sin \left[2\theta (r)\right] |\vec s_1|^2 s_{3,\xi}$, which differs from zero only in the vicinity of the vortex core. Since the magnetostatic modes of a vortex-state dot have zero amplitude at the core center and a small amplitude in its vicinity (except for the modes $m=\pm 1$, for which the amplitudes could be larger due to hybridization with the gyrotropic mode), the core contribution to the three-wave coefficient is weak, and could  become important only in very small dots. In such small dots, however, the SW modes with opposite azimuthal indices are no longer degenerate. Due to the influence of static stray fields of the vortex core, all modes with the same radial index and opposite azimuthal index have different spatial profiles and frequencies, not only the modes with $|m| = 1$ \cite{Ivanov_PRB2002, Taurel_PRB2016}. Also, in a small dot, one may expect three-wave interaction processes involving the gyrotropic mode. The selection rule for such processes result from the fact, that gyrotropic mode exhibits azimuthal dependence characterized by the index $m = \pm 1$ (the sign depends on the vortex core polarity).

Thus, we conclude, that in the case of a vortex-state magnetic dot in a zero external field the three-wave splitting process cannot go into degenerate modes -- the resultant (split) modes should differ either by the modulus of the azimuthal number, or by radial number, or are not frequency-degenerated due to the effect of the vortex core.


\section{Numerical calculations and comparison with experiment and simulations}\label{s:numerical}

In the following, we present results of the numerical calculations of a three-wave interaction efficiency, and the thresholds for the splitting processes in a vortex-state magnetic dot using the above discussed theoretical formalism. The calculations were made for a circular permalloy (Ni$_{81}$Fe$_{19}$ ) dot of the thickness of $h = 50\,\t{nm}$ and diameter $2R = 5.1\,\t{\mu m}$, which was used in the experiment \cite{Schultheiss_PRL2019}. The material parameters of the permalloy are: saturation magnetization $M_s = 810\,\t{kA}/\t{m}$, gyromagnetic ratio $\gamma = 1.86\times10^{11}\,\t{rad}/\t{(sT)}$, exchange constant $A = 1.3\times 10^{−11}\,\t{J}/\t{m}$, and Gilbert damping constant $\alpha_\mathrm{G} = 0.008$.

Analytical or semianalytical theories of the magnetostatic modes of a vortex-state magnetic dot were developed only for the case of a  dominant exchange interaction, when the dipolar interaction can be treated as a perturbation \cite{Ivanov_APL2002, Zivieri_PRB2005}. For our experimental case of a relatively large dot, these theories are not directly applicable for the calculation of the eigenfrequencies and  spatial profiles of the SW modes. Therefore, we used instead a numerical projection method \cite{Buess_PRB2005}. Also, we assumed that in the calculations of both the SW spectrum and the nonlinear coefficients we can safely neglect the effect of the vortex core, since the size of the core is two orders of magnitude smaller than the disk diameter.

The calculated spectrum of the SW modes with radial indices $n = 0,1,2$ is shown in Fig.~\ref{f:1}(b).  Higher-order thickness modes (with thickness index $l = 1,2,3,...$), which were disregarded in the theoretical analysis above,  have frequencies above 10~GHz and, obviously, do not affect dynamics in the studied frequency range.  One can see, that the SW spectrum is rather dense, i.e., the frequency separation between the modes with different azimuthal numbers is relatively small, especially for large $|m|$. Therefore, there exist many different splitting possibilities for any directly excited primary SW mode. Furthermore, this number of possible splitting channels increases even more with the increase of the excitation frequency.

As an example, we consider the splitting of the second radial mode $(2,0)$, which has eigenfrequency of $f_{(2,0)} = \omega_{(2,0)}/(2\pi) = 8.68\,\t{GHz}$, and can be resonantly excited by an out-of-plane microwave magnetic field $b_z$ with a frequency close to the mode eigenfrequency. Analysis of this particular mode (as well as modes with higher radial index) is interesting because, depending on the excitation frequency, the resonance condition for three-wave splitting could be approximately  satisfied for different pairs of split SW modes simultaneously, in our case - pairs of modes having the radial number $n=0,1,2$ (see examples below). In such a case, dependence of three-magnon coefficients affects nonlinear dynamics not only quantitatively, but, also, qualitatively.

\begin{figure}
 \includegraphics[width=0.9\columnwidth]{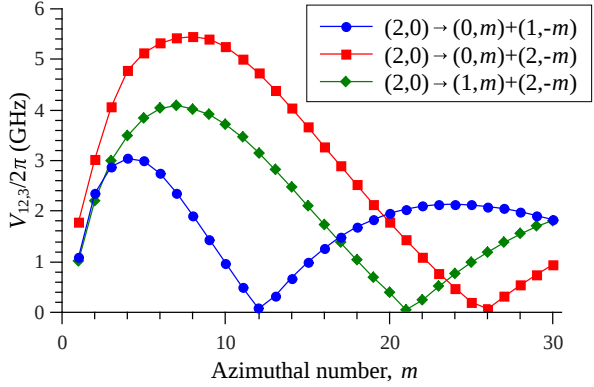}
 \caption{Three-wave interaction efficiency of the radial mode $(2,0)$ with different pairs of azimuthal SW modes.}\label{f:2}
\end{figure}

We performed calculation of the three-wave interaction efficiency for all the possible splitting channels, and the results are presented in Fig.~\ref{f:2}. It is clear, that the three-wave coefficient $V_{12,3}$ demonstrates a significant dependence on both the azimuthal and radial numbers of the split modes. For the split modes having small and moderate azimuthal numbers the largest three-magnon coefficient corresponds to the splitting into $(0,\pm m)$ and $(2,\mp m)$  modes. This could be understood as follows. In an infinite sample three-magnon splitting obeys the rule of momentum (or wave vector) conservation: $\vec k_3 = \vec k_1 + \vec k_2$. Spatial confinement of a finite-size sample breaks this rule. In particular, for a vortex-state dot only the conservation of an angular momentum (i.e. azimuthal index) remains valid. At the same time, the largest three-magnon interaction efficiency is expected for the modes, which approximately satisfy the conservation of the total momentum. Although there is no simple general relation between the total momentum and the mode radial indices, the radial component of the momentum is proportional to the radial index.  Thus, it is natural to expect the maximum interaction efficiency for the modes which satisfy the conservation of the radial index: $n_3 = n_1+n_2$. In our case the channel $(2,0) \to (2,\pm m) + (0,\mp m)$ is the only one which satisfies this relation. At a sufficiently large azimuthal index SW modes become more concentrated at the disk periphery (``whispering gallery magnons''), and the lower is the radial mode number, the more pronounced is this concentration. Consequently, there is a change in the effective volume occupied by the mode, which affects the three-magnon interaction efficiency. As a result, the largest three-magnon coefficient for $m \in [21,30]$ is observed for a different splitting channel -  $n_1 = 0$ and $n_2 = 1$ . With further increase of the index $m$ the most efficient channel is changed many times, but, overall, the interaction efficiency decreases because of the smaller volume occupied by the split modes.

This significant dependence of three-wave interaction efficiency on the azimuthal and radial numbers of the split modes, naturally, should strongly affect the nonlinear dynamics of the SW modes that is realized in experiment. To illustrate this, we set the driving frequency to $f_\t{p} = 8.3\,\t{GHz}$, which efficiently excites the radial mode $(2,0)$ (a little bit off resonance). It is important to stress, that the directly excited mode oscillates, naturally, at the frequency of the driving signal. Therefore, when considering the resonance condition for the three-wave splitting, one should use the driving frequency in the determination of the detuning from the resonance condition, $\delta f = f_\t{p} - (f_1 + f_2)$. Here, $f_{1,2}$ are the eigenfrequencies of the possible split modes which satisfy the above established selection rules.

\begin{figure}
 \includegraphics[width=0.9\columnwidth]{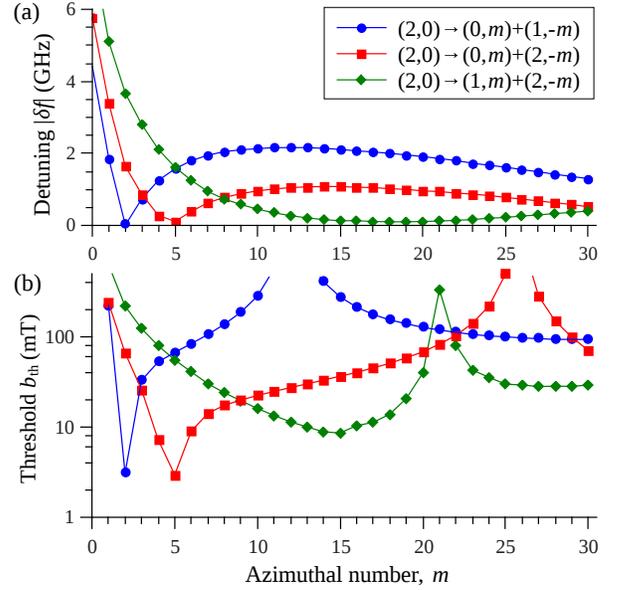}
 \caption{(a) Detuning from the resonance condition of a three-wave splitting,  and (b) the threshold of the splitting of the SW mode $(2,0)$ excited at the frequency of $f_\t{p} = 8.3\,\t{GHz}$ for different splitting channels}\label{f:3}
\end{figure}

The dependence of the detuning $\delta f$ on the azimuthal number of the split modes is shown in Fig.~\ref{f:3}(a) for three possible ($m$-dependent) splitting channels which all satisfy the selection rules. One can find that the resonance condition for the splitting is approximately satisfied for several splitting possibilities, namely $(2,0) \to (0,2) + (1,-2)$, $(2,0) \to (0,5) + (2,-5)$ and $(2,0) \to (1,17) + (2,-17)$, and channels with opposite sign of the azimuthal index, which are degenerate with these ones (e.g., $(2,0) \to (0,-2) + (1,2)$). However, the three-magnon interaction efficiency for these channels differs significantly. The largest efficiency is exhibited by the channel $(2,0) \to (0,5) + (2,-5)$, $V_{12,3}/2\pi = 5.13\,\t{GHz}$, while for other possibilities it is equal to $V_{12,3}/2\pi = 2.35\,\t{GHz}$ and $V_{12,3}/2\pi = 1.38\,\t{GHz}$.

Three-wave splitting is a threshold process, and it starts when the amplitude of the directly excited mode driven by a microwave field, exceed a certain threshold \cite{White_PR1963, Lvov_Book1994}. In the general case of a nonresonant splitting, the threshold for a particular splitting process is determined by the three-wave interaction coefficient, the detuning from the resonance condition, and by the damping rates of the split modes (see Eqs.~(\ref{e:A_th}, \ref{e:b_th}) in the Appendix). As one can see from Fig.~\ref{f:3}(b), in our case, the splitting threshold is the lowest for the splitting into modes $(0,5) + (2,-5)$, which correspond to the largest three-wave  coefficient among the channels close to the resonance condition for the splitting. Therefore, in experiment, this splitting process should be realized.

It is important to note, that even in the case when other channels formally have the thresholds which are close to the lowest one, the realization of these other channel splitting would be practically impossible by a simple increase of the microwave driving field. This happens because as soon as the splitting into the modes having the lowest threshold begins, the nonlinear interaction between the SW modes decreases the ``effective pumping'' for the other modes, and observation of other splitting  channels would  require a much higher amplitude of the driving field than it  formally follows from the calculated thresholds  for  these channels \cite{Lvov_Book1994}.

Our theoretical conclusions are confirmed by the experimental data.  At the excitation frequency of $8.3\,\t{GHz}$, we observed splitting of the mode $(2,0)$ into the modes with azimuthal number $m = \pm 5$ and radial numbers $n_1=0$ and $n_2 = 2$ \cite{Schultheiss_PRL2019}. Note, that the splitting processes into modes $(n_1,m)+(n_2,-m)$  and $(n_1,-m)+(n_2,m)$ are degenerate, and are characterized by the same threshold (except for the small dots in which the effect of the vortex core is relevant). Therefore, splitting into one or the other pair of modes is a random process driven by thermal fluctuations. Hence, in the experiment, we observed standing patterns of azimuthal modes which are a superposition of modes with opposite azimuthal numbers.

\begin{figure}
 \includegraphics[width=0.9\columnwidth]{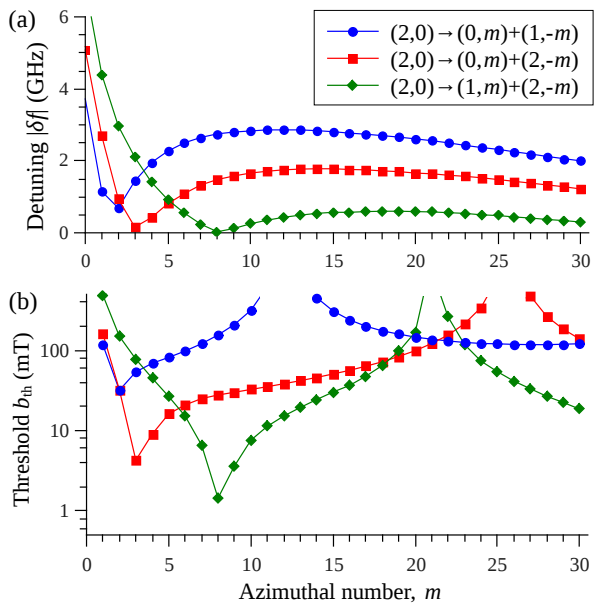}
 \caption{ (a) Detuning from the three-wave splitting resonance condition, and (b) threshold of the splitting of the SW mode $(2,0)$ excited at the frequency of $f_\t{p} = 9\,\t{GHz}$ corresponding to different splitting channels.}\label{f:4}
\end{figure}

Let us now consider how a variation of the excitation frequency affects the splitting process. For this reason, we repeat the same calculations, but for the excitation frequency of $f_\t{p} = 9\,\t{GHz}$. In this case, the directly excited mode still is the  mode $(2,0)$. The resonance condition for the splitting is approximately satisfied for the modes $(0,3) + (2,-3)$ and $(1,8)+(2,-8)$  (Fig.~\ref{f:4}(a)). However, the relation between the three-magnon coefficients and the SW damping rates results in a significantly lower threshold for the splitting $(2,0) \to (1,8)+(2,-8)$ (Fig.~\ref{f:4}(b)), which, in full accordance with the theoretical calculations, was, actually, observed in experiment. The change of the excitation frequency to 8.9~GHz results in the splitting into modes $(1,9)+(2,-9)$. Thus, we can conclude that a variation of the excitation frequency is an efficient way to select the splitting channel. Note, that not only the azimuthal mode number of the split modes can be changed, but also the radial mode numbers.

Experimentally, it is not an easy task to determine the threshold field $b_z^\t{th}$ because the exact power arriving at the microwave antenna is unknown. Thus, to make a quantitative verification of our theoretical calculations, we performed a set of micromagnetic simulations using the MuMax$^3$ software \cite{Vansteenkiste_AIPAdv2014}. Material parameters used in the simulations are the same as mentioned above, the cell size was set to 10$\times$10$\times$50\,nm$^3$. In order to obtain the threshold fields for a given excitation frequency $f_\t{p}$, the microwave field was applied at some field magnitude above threshold and then slowly decreased over the duration of $1\,\t{\mu s}$. The temporal evolution of the split modes was extracted by performing a short-time Fourier transform of the total magnetic energy. The approximate threshold fields were then obtained from the field values at which the secondary modes  disappeared. After this, the mode profiles were obtained by an additional simulation with a fixed microwave field just above threshold.

\begin{table}
\begin{center}
\begin{tabular}{|c|c|c|c|c|}
 \hline
   $f_\t{p}$  & direct & split  & $b_\t{th}$ (mT), & $b_\t{th}$ (mT), \\
   (GHz) &  mode & modes & theory & simulations \\ \hline
  6.1 & $(0,0)$ & $(0,\pm12) + (1,\mp 12)$ & 1.12 & 1.26 \\ \hline
  7.2 & $(1,0)$ & $(0,\pm4) + (1,\mp 4)$ & 2.75 & 3.4 \\ \hline
  8.3 & $(2,0)$ & $(0,\pm5) + (2,\mp 5)$ & 2.95 & 2.62 \\ \hline
  8.9 & $(2,0)$ & $(1,\pm9) + (2,\mp 9)$ & 0.93 & 1.8 \\ \hline
  9.0 & $(2,0)$ & $(1,\pm8) + (2,\mp 8)$ & 1.45 & 2.1 \\ \hline
\end{tabular}
\end{center}
\caption{Channel and threshold of three-magnon splitting at different excitation frequencies. Split modes are the same in theoretical calculations and micromagnetic simulations, as well as coincide with experimental data \cite{Schultheiss_PRL2019}.}\label{t:1}
\end{table}

Results of the micromagnetic simulations are presented in Table~\ref{t:1}, and are compared to the results of the theoretical calculations. For all the considered excitation frequencies, the pair of split modes is the same in theoretical predictions and in micromagnetic simulations, and coincides with the experimental data \cite{Schultheiss_PRL2019}. Theoretically calculated values of the splitting threshold $b_\mathrm{th}$  show a reasonable correspondence to results of the micromagnetic simulations. A mismatch is evident only for the excitation frequency of 8.9~GHz. The observed quantitative discrepancies are, most likely, related to the high sensitivity of the threshold to the SW mode eigenfrequencies (see Eq.~\eqref{e:A_th}), the calculation of which exhibits a limited precision due to the model approximations, finite number of used basis functions  and numerical errors. Nevertheless, our relatively simple calculations provide a good understanding of the process of three-wave splitting of directly excited modes in vortex-state magnetic dots, and allow us to perform  planning of further experiments on the nonlinear magnetization dynamics  in vortex-state magnetic dots.


\section{Summary}\label{s:summary}

In summary, we have investigated the process of three-wave splitting of a directly excited SW mode  in a vortex-state magnetic disk. Using the vector Hamiltonian formalism for nonlinear SW dynamics, we derived expressions for the coefficients of the three-wave interactions between the SW modes of a vortex-state magnetic dot. A qualitative analysis of these expressions yields that three-wave splitting process always obeys one selection rule: conservation of the total azimuthal mode numbers: $m_1+m_2 = m_3$. Additionally, if the directly excited mode is radially symmetric (i.e., $m = 0$), the resultant (split) modes must not have the same radial mode number: $n_1 \neq n_2$. The second rule is relaxed in small magnetic disks due to the influence of the vortex core, which is also responsible for the lifting of the degeneracy of modes with opposite azimuthal indices. Thus, the three-wave splitting in an unbiased (zero external bias magnetic field) vortex-state dot goes always into a pair of frequency non-degenerate SW modes.

The efficiency of the three-wave interactions shows a significant dependence on both the azimuthal and the radial numbers of the split modes. If several split channels are simultaneously close to the fulfillment of the resonance condition for the three-wave splitting, this dependence of the three-magnon coefficients becomes crucial for  the determination of which splitting channel will be, actually, realized in an experiment. The presented theory allows one to predict the splitting channel, and gives a good estimation for the splitting threshold. Furthermore, it opens a possibility for the investigation and quantitative simulation of more complex three-wave scattering processes, e.g., scattering taking place at excitation powers substantially exceeding  the threshold  or the stimulated scattering processes.

\section*{Acknowledgements}

This work was supported in part by the U.S. National Science Foundation (Grants \# EFMA-1641989 and \# ECCS-1708982), by the U.S. Air Force Office of Scientific Research under the MURI grant \# FA9550-19-1-0307, by the Oakland University Foundation, by National Academy of Sciences of Ukraine through the project \#23-04/01-2020 (R.V.), and by the DFG within programs SCHU 2922/1-1 (H.S.) and KA 5069/1-1 (L.K.). K.S. acknowledges funding within the Helmholtz Postdoc Programme.

\section*{Appendix - threshold of nonresonant three-magnon splitting}
\setcounter{equation}{0}
\renewcommand{\theequation}{A\arabic{equation}}

Here, we derive an expression for the three-wave splitting threshold in a general case. In this expression, we take into account (i) a detuning from the resonance condition which appears due to the discreteness of the SW spectrum,  and (ii) different damping rates of the resultant (split) modes, which are expected due to the different frequencies of the the SW modes, and, possibly, different averaged precession ellipticities of the modes. The expressions for the three-wave splitting threshold in particular cases, that take into account either point (i) or (ii) separately, are well-known, and can be found in literature \cite{Lvov_Book1994}.

The threshold for three-wave  splitting can be derived from the dynamic equations for the complex mode amplitudes $c_\nu$, in which it is sufficient to retain only the linear terms \cite{Lvov_Book1994}:
  \begin{equation}
   \begin{split}
    \frac{dc_1}{dt} &+ i\omega_1 c_1 + \Gamma_1 c_1 = i V_{12,3} c_2^* c_3 \,, \\
    \frac{dc_2^*}{dt} & - i\omega_2 c_2^* + \Gamma_2 c_2^* = - i V_{12,3}^* c_1 c_3^* \ .
   \end{split}
  \end{equation}

Here $\omega_{1,2}$ are the eigenfrequencies of the split modes and $\Gamma_{1,2}$ are their damping rates which can be calculated using  SW profiles numerically found within the formalism presented in Ref.~\onlinecite{Verba_PRB2018}. The directly driven SW mode oscillates at the  frequency $\omega_\t{p}$ of the driving microwave field, $c_3 = C_3 e^{-i\omega_\t{p} t}$. The solutions for the split SW modes are searched in the standard form: $c_1 = C_1 e^{-i\tilde\omega_1 t + \alpha t}$, $c_2^* = C_2^* e^{i\tilde\omega_2 t + \alpha t}$. In general, the oscillation frequencies $\tilde\omega_{1,2}$ are unknown and are not equal to the mode eigenfrequencies, but satisfy the relation $\omega_\t{p} = \tilde\omega_1 + \tilde\omega_2$. The parameter $\alpha$ is the growth increment which is negative below the threshold, and positive above the threshold (which means an exponential increase of the split-mode amplitudes from a thermal level). Exactly at the threshold, it is equal to $\alpha = 0$ which yields the following characteristic equation:
  \begin{equation}
   [-i(\tilde\omega_1 - \omega_1) + \Gamma_1][i(\tilde\omega_2 - \omega_2) + \Gamma_2] = |V_{12,3}C_3|^2 \ .
  \end{equation}
From the requirement of a zero imaginary part of the left-hand part (as it stands for the right-hand one), one finds the relation between the oscillation frequencies of split modes: $(\tilde\omega_1 - \omega_1)/\Gamma_1 = (\tilde\omega_2 - \omega_2)/\Gamma_2$, using which the threshold value of the directly driven SW mode amplitude is found in the form:
  \begin{equation}\label{e:A_th}
   |V_{12,3}C_3|^2 = \Gamma_1 \Gamma_2 \left(1+ \frac{\delta\omega^2}{(\Gamma_1+\Gamma_2)^2} \right) \,,
  \end{equation}
where $\delta\omega = \omega_\t{p} - (\omega_1+ \omega_2)$ is the detuning from the resonance condition. The amplitude of the directly excited mode is related to the driving microwave field by the usual expression:
  \begin{equation}\label{e:b_th}
   C_3 = \frac{\gamma b_z \langle s_{3,z} \rangle}{2\sqrt{\left(\omega_\t{p} - \omega_3 \right)^2 + \Gamma_3^2}} \,,
  \end{equation}
where $\langle s_z \rangle$ is the averaged out-of-plane dynamic component of the SW mode $\vec s_3$.


\begin{thebibliography}{47}%
\makeatletter
\providecommand \@ifxundefined [1]{%
 \@ifx{#1\undefined}
}%
\providecommand \@ifnum [1]{%
 \ifnum #1\expandafter \@firstoftwo
 \else \expandafter \@secondoftwo
 \fi
}%
\providecommand \@ifx [1]{%
 \ifx #1\expandafter \@firstoftwo
 \else \expandafter \@secondoftwo
 \fi
}%
\providecommand \natexlab [1]{#1}%
\providecommand \enquote  [1]{``#1''}%
\providecommand \bibnamefont  [1]{#1}%
\providecommand \bibfnamefont [1]{#1}%
\providecommand \citenamefont [1]{#1}%
\providecommand \href@noop [0]{\@secondoftwo}%
\providecommand \href [0]{\begingroup \@sanitize@url \@href}%
\providecommand \@href[1]{\@@startlink{#1}\@@href}%
\providecommand \@@href[1]{\endgroup#1\@@endlink}%
\providecommand \@sanitize@url [0]{\catcode `\\12\catcode `\$12\catcode
  `\&12\catcode `\#12\catcode `\^12\catcode `\_12\catcode `\%12\relax}%
\providecommand \@@startlink[1]{}%
\providecommand \@@endlink[0]{}%
\providecommand \url  [0]{\begingroup\@sanitize@url \@url }%
\providecommand \@url [1]{\endgroup\@href {#1}{\urlprefix }}%
\providecommand \urlprefix  [0]{URL }%
\providecommand \Eprint [0]{\href }%
\providecommand \doibase [0]{http://dx.doi.org/}%
\providecommand \selectlanguage [0]{\@gobble}%
\providecommand \bibinfo  [0]{\@secondoftwo}%
\providecommand \bibfield  [0]{\@secondoftwo}%
\providecommand \translation [1]{[#1]}%
\providecommand \BibitemOpen [0]{}%
\providecommand \bibitemStop [0]{}%
\providecommand \bibitemNoStop [0]{.\EOS\space}%
\providecommand \EOS [0]{\spacefactor3000\relax}%
\providecommand \BibitemShut  [1]{\csname bibitem#1\endcsname}%
\let\auto@bib@innerbib\@empty
\bibitem [{\citenamefont {Kosevich}\ \emph {et~al.}(1983)\citenamefont
  {Kosevich}, \citenamefont {Ivanov},\ and\ \citenamefont
  {Kovalev}}]{Kosevich_Book}%
  \BibitemOpen
  \bibfield  {author} {\bibinfo {author} {\bibfnamefont {A.~M.}\ \bibnamefont
  {Kosevich}}, \bibinfo {author} {\bibfnamefont {B.~A.}\ \bibnamefont
  {Ivanov}}, \ and\ \bibinfo {author} {\bibfnamefont {A.~S.}\ \bibnamefont
  {Kovalev}},\ }\href@noop {} {\emph {\bibinfo {title} {{Nonlinear
  Magentization Waves. Dynamic and Topological Solitons}}}}\ (\bibinfo
  {publisher} {Naukova dumka, Kyiv},\ \bibinfo {year} {1983})\ \bibinfo {note}
  {in russian}\BibitemShut {NoStop}%
\bibitem [{\citenamefont {Gurevich}\ and\ \citenamefont
  {Melkov}(1996)}]{Gurevich_Book1996}%
  \BibitemOpen
  \bibfield  {author} {\bibinfo {author} {\bibfnamefont {A.~G.}\ \bibnamefont
  {Gurevich}}\ and\ \bibinfo {author} {\bibfnamefont {G.~A.}\ \bibnamefont
  {Melkov}},\ }\href@noop {} {\emph {\bibinfo {title} {{Magnetization
  Oscillations and Waves}}}}\ (\bibinfo  {publisher} {CRC Press, New York},\
  \bibinfo {year} {1996})\ p.\ \bibinfo {pages} {464}\BibitemShut {NoStop}%
\bibitem [{\citenamefont {Bertotti}\ \emph {et~al.}(2009)\citenamefont
  {Bertotti}, \citenamefont {Mayergoyz},\ and\ \citenamefont
  {Serpico}}]{Bertotti_Book}%
  \BibitemOpen
  \bibfield  {author} {\bibinfo {author} {\bibfnamefont {G.}~\bibnamefont
  {Bertotti}}, \bibinfo {author} {\bibfnamefont {I.}~\bibnamefont {Mayergoyz}},
  \ and\ \bibinfo {author} {\bibfnamefont {C.}~\bibnamefont {Serpico}},\
  }\href@noop {} {\emph {\bibinfo {title} {{Nonlinear Magnetization Dynamics in
  Nanosystems}}}}\ (\bibinfo  {publisher} {Elsevier},\ \bibinfo {year}
  {2009})\BibitemShut {NoStop}%
\bibitem [{\citenamefont {Schl{\"o}mann}(1959)}]{Schlomann_PR1959}%
  \BibitemOpen
  \bibfield  {author} {\bibinfo {author} {\bibfnamefont {E.}~\bibnamefont
  {Schl{\"o}mann}},\ }\href {\doibase 10.1103/PhysRev.116.828} {\bibfield
  {journal} {\bibinfo  {journal} {Phys. Rev.}\ }\textbf {\bibinfo {volume}
  {116}},\ \bibinfo {pages} {828} (\bibinfo {year} {1959})}\BibitemShut
  {NoStop}%
\bibitem [{\citenamefont {L'vov}(1994)}]{Lvov_Book1994}%
  \BibitemOpen
  \bibfield  {author} {\bibinfo {author} {\bibfnamefont {V.~S.}\ \bibnamefont
  {L'vov}},\ }\href@noop {} {\emph {\bibinfo {title} {{Wave Turbulence under
  Parametric Excitation}}}}\ (\bibinfo  {publisher} {Springer-Verlag, New
  York},\ \bibinfo {year} {1994})\BibitemShut {NoStop}%
\bibitem [{\citenamefont {Safonov}(2013)}]{Safonov_Book}%
  \BibitemOpen
  \bibfield  {author} {\bibinfo {author} {\bibfnamefont {V.~L.}\ \bibnamefont
  {Safonov}},\ }\href@noop {} {\emph {\bibinfo {title} {{Noneequilibrium
  Magnons:Theory, Experiment and Applications}}}}\ (\bibinfo  {publisher}
  {Wiley-VCH, Germany},\ \bibinfo {year} {2013})\BibitemShut {NoStop}%
\bibitem [{\citenamefont {Gottlieb}\ and\ \citenamefont
  {Suhl}(1962)}]{Gottlieb_JAP1962}%
  \BibitemOpen
  \bibfield  {author} {\bibinfo {author} {\bibfnamefont {P.}~\bibnamefont
  {Gottlieb}}\ and\ \bibinfo {author} {\bibfnamefont {H.}~\bibnamefont
  {Suhl}},\ }\href {\doibase 10.1063/1.1728762} {\bibfield  {journal} {\bibinfo
   {journal} {J. Appl. Phys.}\ }\textbf {\bibinfo {volume} {33}},\ \bibinfo
  {pages} {1508} (\bibinfo {year} {1962})}\BibitemShut {NoStop}%
\bibitem [{\citenamefont {White}\ and\ \citenamefont
  {Sparks}(1963)}]{White_PR1963}%
  \BibitemOpen
  \bibfield  {author} {\bibinfo {author} {\bibfnamefont {R.~M.}\ \bibnamefont
  {White}}\ and\ \bibinfo {author} {\bibfnamefont {M.}~\bibnamefont {Sparks}},\
  }\href {\doibase 10.1103/PhysRev.130.632} {\bibfield  {journal} {\bibinfo
  {journal} {Phys. Rev.}\ }\textbf {\bibinfo {volume} {130}},\ \bibinfo {pages}
  {632} (\bibinfo {year} {1963})}\BibitemShut {NoStop}%
\bibitem [{\citenamefont {Patton}(1979)}]{Patton_PSS1979}%
  \BibitemOpen
  \bibfield  {author} {\bibinfo {author} {\bibfnamefont {C.~E.}\ \bibnamefont
  {Patton}},\ }\href {\doibase 10.1002/pssb.2220920124} {\bibfield  {journal}
  {\bibinfo  {journal} {Phys. Status Solidi (b)}\ }\textbf {\bibinfo {volume}
  {92}},\ \bibinfo {pages} {211} (\bibinfo {year} {1979})}\BibitemShut
  {NoStop}%
\bibitem [{\citenamefont {Nazarov}\ \emph {et~al.}(2002)\citenamefont
  {Nazarov}, \citenamefont {Patton}, \citenamefont {Cox}, \citenamefont
  {Chen},\ and\ \citenamefont {Kabos}}]{Nazarov_JMMM2002}%
  \BibitemOpen
  \bibfield  {author} {\bibinfo {author} {\bibfnamefont {A.}~\bibnamefont
  {Nazarov}}, \bibinfo {author} {\bibfnamefont {C.}~\bibnamefont {Patton}},
  \bibinfo {author} {\bibfnamefont {R.}~\bibnamefont {Cox}}, \bibinfo {author}
  {\bibfnamefont {L.}~\bibnamefont {Chen}}, \ and\ \bibinfo {author}
  {\bibfnamefont {P.}~\bibnamefont {Kabos}},\ }\href {\doibase
  10.1016/S0304-8853(02)00171-3} {\bibfield  {journal} {\bibinfo  {journal} {J.
  Magn. Magn. Mater.}\ }\textbf {\bibinfo {volume} {248}},\ \bibinfo {pages}
  {164} (\bibinfo {year} {2002})}\BibitemShut {NoStop}%
\bibitem [{\citenamefont {Boardman}\ and\ \citenamefont
  {Nikitov}(1988)}]{Boardman_PRB1988}%
  \BibitemOpen
  \bibfield  {author} {\bibinfo {author} {\bibfnamefont {A.~D.}\ \bibnamefont
  {Boardman}}\ and\ \bibinfo {author} {\bibfnamefont {S.~A.}\ \bibnamefont
  {Nikitov}},\ }\href {\doibase 10.1103/PhysRevB.38.11444} {\bibfield
  {journal} {\bibinfo  {journal} {Phys. Rev. B}\ }\textbf {\bibinfo {volume}
  {38}},\ \bibinfo {pages} {11444} (\bibinfo {year} {1988})}\BibitemShut
  {NoStop}%
\bibitem [{\citenamefont {Liu}\ \emph {et~al.}(2019)\citenamefont {Liu},
  \citenamefont {Riley}, \citenamefont {Ord{\'o}{\~n}ez-Romero}, \citenamefont
  {Kalinikos},\ and\ \citenamefont {Buchanan}}]{Liu_PRB2019}%
  \BibitemOpen
  \bibfield  {author} {\bibinfo {author} {\bibfnamefont {H.~J.~J.}\
  \bibnamefont {Liu}}, \bibinfo {author} {\bibfnamefont {G.~A.}\ \bibnamefont
  {Riley}}, \bibinfo {author} {\bibfnamefont {C.~L.}\ \bibnamefont
  {Ord{\'o}{\~n}ez-Romero}}, \bibinfo {author} {\bibfnamefont {B.~A.}\
  \bibnamefont {Kalinikos}}, \ and\ \bibinfo {author} {\bibfnamefont {K.~S.}\
  \bibnamefont {Buchanan}},\ }\href {\doibase 10.1103/PhysRevB.99.024429}
  {\bibfield  {journal} {\bibinfo  {journal} {Phys. Rev. B}\ }\textbf {\bibinfo
  {volume} {99}},\ \bibinfo {pages} {024429} (\bibinfo {year}
  {2019})}\BibitemShut {NoStop}%
\bibitem [{\citenamefont {Suhl}(1960)}]{Suhl_JAP1960}%
  \BibitemOpen
  \bibfield  {author} {\bibinfo {author} {\bibfnamefont {H.}~\bibnamefont
  {Suhl}},\ }\href {\doibase 10.1063/1.1735723} {\bibfield  {journal} {\bibinfo
   {journal} {J. Appl. Phys.}\ }\textbf {\bibinfo {volume} {31}},\ \bibinfo
  {pages} {935} (\bibinfo {year} {1960})}\BibitemShut {NoStop}%
\bibitem [{\citenamefont {Gui}\ \emph {et~al.}(2009)\citenamefont {Gui},
  \citenamefont {Wirthmann}, \citenamefont {Mecking},\ and\ \citenamefont
  {Hu}}]{Gui_PRB2009}%
  \BibitemOpen
  \bibfield  {author} {\bibinfo {author} {\bibfnamefont {Y.~S.}\ \bibnamefont
  {Gui}}, \bibinfo {author} {\bibfnamefont {A.}~\bibnamefont {Wirthmann}},
  \bibinfo {author} {\bibfnamefont {N.}~\bibnamefont {Mecking}}, \ and\
  \bibinfo {author} {\bibfnamefont {C.-M.}\ \bibnamefont {Hu}},\ }\href
  {\doibase 10.1103/PhysRevB.80.060402} {\bibfield  {journal} {\bibinfo
  {journal} {Phys. Rev. B}\ }\textbf {\bibinfo {volume} {80}},\ \bibinfo
  {pages} {060402(R)} (\bibinfo {year} {2009})}\BibitemShut {NoStop}%
\bibitem [{\citenamefont {Zakharov}\ \emph {et~al.}(1975)\citenamefont
  {Zakharov}, \citenamefont {L'vov},\ and\ \citenamefont
  {Starobinets}}]{Zakharov_UFN1975}%
  \BibitemOpen
  \bibfield  {author} {\bibinfo {author} {\bibfnamefont {V.~E.}\ \bibnamefont
  {Zakharov}}, \bibinfo {author} {\bibfnamefont {V.~S.}\ \bibnamefont {L'vov}},
  \ and\ \bibinfo {author} {\bibfnamefont {S.~S.}\ \bibnamefont
  {Starobinets}},\ }\href {\doibase 10.1070/PU1975v017n06ABEH004404} {\bibfield
   {journal} {\bibinfo  {journal} {Sov. Phys. Usp.}\ }\textbf {\bibinfo
  {volume} {17}},\ \bibinfo {pages} {896} (\bibinfo {year} {1975})}\BibitemShut
  {NoStop}%
\bibitem [{\citenamefont {Kalinikos}\ \emph {et~al.}(1988)\citenamefont
  {Kalinikos}, \citenamefont {Kovshikov},\ and\ \citenamefont
  {Slavin}}]{Kalinikos_JETP1988}%
  \BibitemOpen
  \bibfield  {author} {\bibinfo {author} {\bibfnamefont {B.~A.}\ \bibnamefont
  {Kalinikos}}, \bibinfo {author} {\bibfnamefont {N.~G.}\ \bibnamefont
  {Kovshikov}}, \ and\ \bibinfo {author} {\bibfnamefont {A.~N.}\ \bibnamefont
  {Slavin}},\ }\href@noop {} {\bibfield  {journal} {\bibinfo  {journal} {Sov.
  Phys. JETP}\ }\textbf {\bibinfo {volume} {67}},\ \bibinfo {pages} {303}
  (\bibinfo {year} {1988})}\BibitemShut {NoStop}%
\bibitem [{\citenamefont {Chen}\ \emph {et~al.}(1993)\citenamefont {Chen},
  \citenamefont {Tsankov}, \citenamefont {Nash},\ and\ \citenamefont
  {Patton}}]{Chen_PRL1993}%
  \BibitemOpen
  \bibfield  {author} {\bibinfo {author} {\bibfnamefont {M.}~\bibnamefont
  {Chen}}, \bibinfo {author} {\bibfnamefont {M.~A.}\ \bibnamefont {Tsankov}},
  \bibinfo {author} {\bibfnamefont {J.~M.}\ \bibnamefont {Nash}}, \ and\
  \bibinfo {author} {\bibfnamefont {C.~E.}\ \bibnamefont {Patton}},\ }\href
  {\doibase 10.1103/PhysRevLett.70.1707} {\bibfield  {journal} {\bibinfo
  {journal} {Phys. Rev. Lett.}\ }\textbf {\bibinfo {volume} {70}},\ \bibinfo
  {pages} {1707} (\bibinfo {year} {1993})}\BibitemShut {NoStop}%
\bibitem [{\citenamefont {Melkov}\ \emph {et~al.}(2001)\citenamefont {Melkov},
  \citenamefont {Kobljanskyj}, \citenamefont {Serga}, \citenamefont
  {Tiberkevich},\ and\ \citenamefont {Slavin}}]{Melkov_JAP2001}%
  \BibitemOpen
  \bibfield  {author} {\bibinfo {author} {\bibfnamefont {G.~A.}\ \bibnamefont
  {Melkov}}, \bibinfo {author} {\bibfnamefont {Y.~V.}\ \bibnamefont
  {Kobljanskyj}}, \bibinfo {author} {\bibfnamefont {A.~A.}\ \bibnamefont
  {Serga}}, \bibinfo {author} {\bibfnamefont {V.~S.}\ \bibnamefont
  {Tiberkevich}}, \ and\ \bibinfo {author} {\bibfnamefont {A.~N.}\ \bibnamefont
  {Slavin}},\ }\href {\doibase 10.1063/1.1357141} {\bibfield  {journal}
  {\bibinfo  {journal} {J. Appl. Phys.}\ }\textbf {\bibinfo {volume} {89}},\
  \bibinfo {pages} {6689} (\bibinfo {year} {2001})}\BibitemShut {NoStop}%
\bibitem [{\citenamefont {Livesey}\ \emph {et~al.}(2007)\citenamefont
  {Livesey}, \citenamefont {Kostylev},\ and\ \citenamefont
  {Stamps}}]{Livesey_PRB2007}%
  \BibitemOpen
  \bibfield  {author} {\bibinfo {author} {\bibfnamefont {K.~L.}\ \bibnamefont
  {Livesey}}, \bibinfo {author} {\bibfnamefont {M.~P.}\ \bibnamefont
  {Kostylev}}, \ and\ \bibinfo {author} {\bibfnamefont {R.~L.}\ \bibnamefont
  {Stamps}},\ }\href {\doibase 10.1103/PhysRevB.75.174427} {\bibfield
  {journal} {\bibinfo  {journal} {Phys. Rev. B}\ }\textbf {\bibinfo {volume}
  {75}},\ \bibinfo {pages} {174427} (\bibinfo {year} {2007})}\BibitemShut
  {NoStop}%
\bibitem [{\citenamefont {Melkov}\ \emph {et~al.}(2013)\citenamefont {Melkov},
  \citenamefont {Slobodianiuk}, \citenamefont {Tiberkevich}, \citenamefont
  {de~Loubens}, \citenamefont {Klein},\ and\ \citenamefont
  {Slavin}}]{Melkov_MagLet2013}%
  \BibitemOpen
  \bibfield  {author} {\bibinfo {author} {\bibfnamefont {G.~A.}\ \bibnamefont
  {Melkov}}, \bibinfo {author} {\bibfnamefont {D.~V.}\ \bibnamefont
  {Slobodianiuk}}, \bibinfo {author} {\bibfnamefont {V.~S.}\ \bibnamefont
  {Tiberkevich}}, \bibinfo {author} {\bibfnamefont {G.}~\bibnamefont
  {de~Loubens}}, \bibinfo {author} {\bibfnamefont {O.}~\bibnamefont {Klein}}, \
  and\ \bibinfo {author} {\bibfnamefont {A.~N.}\ \bibnamefont {Slavin}},\
  }\href {\doibase 10.1109/LMAG.2013.2278682} {\bibfield  {journal} {\bibinfo
  {journal} {IEEE Magn. Lett.}\ }\textbf {\bibinfo {volume} {4}},\ \bibinfo
  {pages} {4000504} (\bibinfo {year} {2013})}\BibitemShut {NoStop}%
\bibitem [{\citenamefont {{Slobodianiuk}}\ \emph {et~al.}(2019)\citenamefont
  {{Slobodianiuk}}, \citenamefont {{Melkov}}, \citenamefont {{Schultheiss}},
  \citenamefont {{Schultheiss}},\ and\ \citenamefont
  {{Verba}}}]{Slobodianiuk_MagLett2019}%
  \BibitemOpen
  \bibfield  {author} {\bibinfo {author} {\bibfnamefont {D.~V.}\ \bibnamefont
  {{Slobodianiuk}}}, \bibinfo {author} {\bibfnamefont {G.~A.}\ \bibnamefont
  {{Melkov}}}, \bibinfo {author} {\bibfnamefont {K.}~\bibnamefont
  {{Schultheiss}}}, \bibinfo {author} {\bibfnamefont {H.}~\bibnamefont
  {{Schultheiss}}}, \ and\ \bibinfo {author} {\bibfnamefont {R.~V.}\
  \bibnamefont {{Verba}}},\ }\href {\doibase 10.1109/LMAG.2019.2913132}
  {\bibfield  {journal} {\bibinfo  {journal} {IEEE Magn. Lett.}\ }\textbf
  {\bibinfo {volume} {10}},\ \bibinfo {pages} {6103405} (\bibinfo {year}
  {2019})}\BibitemShut {NoStop}%
\bibitem [{\citenamefont {Barsukov}\ \emph {et~al.}(2019)\citenamefont
  {Barsukov}, \citenamefont {Lee}, \citenamefont {Jara}, \citenamefont {Chen},
  \citenamefont {Gon\c{c}alves}, \citenamefont {Sha}, \citenamefont {Katine},
  \citenamefont {Arias}, \citenamefont {Ivanov},\ and\ \citenamefont
  {Krivorotov}}]{Barsukov_SciAdv2019}%
  \BibitemOpen
  \bibfield  {author} {\bibinfo {author} {\bibfnamefont {I.}~\bibnamefont
  {Barsukov}}, \bibinfo {author} {\bibfnamefont {H.~K.}\ \bibnamefont {Lee}},
  \bibinfo {author} {\bibfnamefont {A.~A.}\ \bibnamefont {Jara}}, \bibinfo
  {author} {\bibfnamefont {Y.-J.}\ \bibnamefont {Chen}}, \bibinfo {author}
  {\bibfnamefont {A.~M.}\ \bibnamefont {Gon\c{c}alves}}, \bibinfo {author}
  {\bibfnamefont {C.}~\bibnamefont {Sha}}, \bibinfo {author} {\bibfnamefont
  {J.~A.}\ \bibnamefont {Katine}}, \bibinfo {author} {\bibfnamefont {R.~E.}\
  \bibnamefont {Arias}}, \bibinfo {author} {\bibfnamefont {B.~A.}\ \bibnamefont
  {Ivanov}}, \ and\ \bibinfo {author} {\bibfnamefont {I.~N.}\ \bibnamefont
  {Krivorotov}},\ }\href {\doibase 10.1126/sciadv.aav6943} {\bibfield
  {journal} {\bibinfo  {journal} {Sci. Adv.}\ }\textbf {\bibinfo {volume}
  {5}},\ \bibinfo {pages} {eaav6943} (\bibinfo {year} {2019})}\BibitemShut
  {NoStop}%
\bibitem [{\citenamefont {Kobljanskyj}\ \emph {et~al.}(2012)\citenamefont
  {Kobljanskyj}, \citenamefont {Melkov}, \citenamefont {Guslienko},
  \citenamefont {Novosad}, \citenamefont {Bader}, \citenamefont {Kostylev},\
  and\ \citenamefont {Slavin}}]{Kobljanskyj_SR2012}%
  \BibitemOpen
  \bibfield  {author} {\bibinfo {author} {\bibfnamefont {Y.}~\bibnamefont
  {Kobljanskyj}}, \bibinfo {author} {\bibfnamefont {G.}~\bibnamefont {Melkov}},
  \bibinfo {author} {\bibfnamefont {K.}~\bibnamefont {Guslienko}}, \bibinfo
  {author} {\bibfnamefont {V.}~\bibnamefont {Novosad}}, \bibinfo {author}
  {\bibfnamefont {S.~D.}\ \bibnamefont {Bader}}, \bibinfo {author}
  {\bibfnamefont {M.}~\bibnamefont {Kostylev}}, \ and\ \bibinfo {author}
  {\bibfnamefont {A.}~\bibnamefont {Slavin}},\ }\href {\doibase
  10.1038/srep00478} {\bibfield  {journal} {\bibinfo  {journal} {Sci. Rep.}\
  }\textbf {\bibinfo {volume} {2}},\ \bibinfo {pages} {478} (\bibinfo {year}
  {2012})}\BibitemShut {NoStop}%
\bibitem [{\citenamefont {Camley}(2014)}]{Camley_PRB2014}%
  \BibitemOpen
  \bibfield  {author} {\bibinfo {author} {\bibfnamefont {R.~E.}\ \bibnamefont
  {Camley}},\ }\href {\doibase 10.1103/PhysRevB.89.214402} {\bibfield
  {journal} {\bibinfo  {journal} {Phys. Rev. B}\ }\textbf {\bibinfo {volume}
  {89}},\ \bibinfo {pages} {214402} (\bibinfo {year} {2014})}\BibitemShut
  {NoStop}%
\bibitem [{\citenamefont {Schultheiss}\ \emph {et~al.}(2019)\citenamefont
  {Schultheiss}, \citenamefont {Verba}, \citenamefont {Wehrmann}, \citenamefont
  {Wagner}, \citenamefont {K{\"o}rber}, \citenamefont {Hula}, \citenamefont
  {Hache}, \citenamefont {K{\'a}kay}, \citenamefont {Awad}, \citenamefont
  {Tiberkevich}, \citenamefont {Slavin}, \citenamefont {Fassbender},\ and\
  \citenamefont {Schultheiss}}]{Schultheiss_PRL2019}%
  \BibitemOpen
  \bibfield  {author} {\bibinfo {author} {\bibfnamefont {K.}~\bibnamefont
  {Schultheiss}}, \bibinfo {author} {\bibfnamefont {R.}~\bibnamefont {Verba}},
  \bibinfo {author} {\bibfnamefont {F.}~\bibnamefont {Wehrmann}}, \bibinfo
  {author} {\bibfnamefont {K.}~\bibnamefont {Wagner}}, \bibinfo {author}
  {\bibfnamefont {L.}~\bibnamefont {K{\"o}rber}}, \bibinfo {author}
  {\bibfnamefont {T.}~\bibnamefont {Hula}}, \bibinfo {author} {\bibfnamefont
  {T.}~\bibnamefont {Hache}}, \bibinfo {author} {\bibfnamefont
  {A.}~\bibnamefont {K{\'a}kay}}, \bibinfo {author} {\bibfnamefont {A.~A.}\
  \bibnamefont {Awad}}, \bibinfo {author} {\bibfnamefont {V.}~\bibnamefont
  {Tiberkevich}}, \bibinfo {author} {\bibfnamefont {A.~N.}\ \bibnamefont
  {Slavin}}, \bibinfo {author} {\bibfnamefont {J.}~\bibnamefont {Fassbender}},
  \ and\ \bibinfo {author} {\bibfnamefont {H.}~\bibnamefont {Schultheiss}},\
  }\href {\doibase 10.1103/PhysRevLett.122.097202} {\bibfield  {journal}
  {\bibinfo  {journal} {Phys. Rev. Lett.}\ }\textbf {\bibinfo {volume} {122}},\
  \bibinfo {pages} {097202} (\bibinfo {year} {2019})}\BibitemShut {NoStop}%
\bibitem [{\citenamefont {{Lord Rayleigh, O.M.,
  F.R.S.}}(1914)}]{Reyleigh_1914}%
  \BibitemOpen
  \bibfield  {author} {\bibinfo {author} {\bibnamefont {{Lord Rayleigh, O.M.,
  F.R.S.}}},\ }\href {\doibase 10.1080/14786440108635067} {\bibfield  {journal}
  {\bibinfo  {journal} {Philos. Mag.}\ }\textbf {\bibinfo {volume} {27}},\
  \bibinfo {pages} {100} (\bibinfo {year} {1914})}\BibitemShut {NoStop}%
\bibitem [{\citenamefont {Tyberkevych}\ \emph {et~al.}()\citenamefont
  {Tyberkevych}, \citenamefont {Slavin}, \citenamefont {Artemchuk},\ and\
  \citenamefont {Rowlands}}]{Tyberkevych_ArXiv}%
  \BibitemOpen
  \bibfield  {author} {\bibinfo {author} {\bibfnamefont {V.}~\bibnamefont
  {Tyberkevych}}, \bibinfo {author} {\bibfnamefont {A.}~\bibnamefont {Slavin}},
  \bibinfo {author} {\bibfnamefont {P.}~\bibnamefont {Artemchuk}}, \ and\
  \bibinfo {author} {\bibfnamefont {G.}~\bibnamefont {Rowlands}},\ }\href@noop
  {} {\enquote {\bibinfo {title} {{Vector Hamiltonian Formalism for Nonlinear
  Magnetization Dynamics}},}\ }\bibinfo {note} {ArXiv:2011.13562
  [cond-mat.mtrl-sci]}\BibitemShut {NoStop}%
\bibitem [{\citenamefont {Dzyapko}\ \emph {et~al.}(2017)\citenamefont
  {Dzyapko}, \citenamefont {Lisenkov}, \citenamefont {Nowik-Boltyk},
  \citenamefont {Demidov}, \citenamefont {Demokritov}, \citenamefont {Koene},
  \citenamefont {Kirilyuk}, \citenamefont {Rasing}, \citenamefont
  {Tiberkevich},\ and\ \citenamefont {Slavin}}]{Dzyapko_PRB2017}%
  \BibitemOpen
  \bibfield  {author} {\bibinfo {author} {\bibfnamefont {O.}~\bibnamefont
  {Dzyapko}}, \bibinfo {author} {\bibfnamefont {I.}~\bibnamefont {Lisenkov}},
  \bibinfo {author} {\bibfnamefont {P.}~\bibnamefont {Nowik-Boltyk}}, \bibinfo
  {author} {\bibfnamefont {V.~E.}\ \bibnamefont {Demidov}}, \bibinfo {author}
  {\bibfnamefont {S.~O.}\ \bibnamefont {Demokritov}}, \bibinfo {author}
  {\bibfnamefont {B.}~\bibnamefont {Koene}}, \bibinfo {author} {\bibfnamefont
  {A.}~\bibnamefont {Kirilyuk}}, \bibinfo {author} {\bibfnamefont
  {T.}~\bibnamefont {Rasing}}, \bibinfo {author} {\bibfnamefont
  {V.}~\bibnamefont {Tiberkevich}}, \ and\ \bibinfo {author} {\bibfnamefont
  {A.}~\bibnamefont {Slavin}},\ }\href {\doibase 10.1103/PhysRevB.96.064438}
  {\bibfield  {journal} {\bibinfo  {journal} {Phys. Rev. B}\ }\textbf {\bibinfo
  {volume} {96}},\ \bibinfo {pages} {064438} (\bibinfo {year}
  {2017})}\BibitemShut {NoStop}%
\bibitem [{\citenamefont {Verba}\ \emph {et~al.}(2016)\citenamefont {Verba},
  \citenamefont {Carpentieri}, \citenamefont {Finocchio}, \citenamefont
  {Tiberkevich},\ and\ \citenamefont {Slavin}}]{Verba_SciRep2016}%
  \BibitemOpen
  \bibfield  {author} {\bibinfo {author} {\bibfnamefont {R.}~\bibnamefont
  {Verba}}, \bibinfo {author} {\bibfnamefont {M.}~\bibnamefont {Carpentieri}},
  \bibinfo {author} {\bibfnamefont {G.}~\bibnamefont {Finocchio}}, \bibinfo
  {author} {\bibfnamefont {V.}~\bibnamefont {Tiberkevich}}, \ and\ \bibinfo
  {author} {\bibfnamefont {A.}~\bibnamefont {Slavin}},\ }\href {\doibase
  10.1038/srep25018} {\bibfield  {journal} {\bibinfo  {journal} {Sci. Rep.}\
  }\textbf {\bibinfo {volume} {6}},\ \bibinfo {pages} {25018} (\bibinfo {year}
  {2016})}\BibitemShut {NoStop}%
\bibitem [{\citenamefont {Krivosik}\ and\ \citenamefont
  {Patton}(2010)}]{Krivosik_PRB2010}%
  \BibitemOpen
  \bibfield  {author} {\bibinfo {author} {\bibfnamefont {P.}~\bibnamefont
  {Krivosik}}\ and\ \bibinfo {author} {\bibfnamefont {C.~E.}\ \bibnamefont
  {Patton}},\ }\href {\doibase 10.1103/PhysRevB.82.184428} {\bibfield
  {journal} {\bibinfo  {journal} {Phys. Rev. B}\ }\textbf {\bibinfo {volume}
  {82}},\ \bibinfo {pages} {184428} (\bibinfo {year} {2010})}\BibitemShut
  {NoStop}%
\bibitem [{\citenamefont {Verba}\ \emph {et~al.}(2019)\citenamefont {Verba},
  \citenamefont {Tiberkevich},\ and\ \citenamefont {Slavin}}]{Verba_PRB2019}%
  \BibitemOpen
  \bibfield  {author} {\bibinfo {author} {\bibfnamefont {R.}~\bibnamefont
  {Verba}}, \bibinfo {author} {\bibfnamefont {V.}~\bibnamefont {Tiberkevich}},
  \ and\ \bibinfo {author} {\bibfnamefont {A.}~\bibnamefont {Slavin}},\ }\href
  {\doibase 10.1103/PhysRevB.99.174431} {\bibfield  {journal} {\bibinfo
  {journal} {Phys. Rev. B}\ }\textbf {\bibinfo {volume} {99}},\ \bibinfo
  {pages} {174431} (\bibinfo {year} {2019})}\BibitemShut {NoStop}%
\bibitem [{\citenamefont {Abyzov}\ and\ \citenamefont
  {Ivanov}(1979)}]{Abyzov_JETP1979}%
  \BibitemOpen
  \bibfield  {author} {\bibinfo {author} {\bibfnamefont {A.~S.}\ \bibnamefont
  {Abyzov}}\ and\ \bibinfo {author} {\bibfnamefont {B.~A.}\ \bibnamefont
  {Ivanov}},\ }\href@noop {} {\bibfield  {journal} {\bibinfo  {journal} {Sov.
  Phys. JETP}\ }\textbf {\bibinfo {volume} {49}},\ \bibinfo {pages} {865}
  (\bibinfo {year} {1979})},\ \bibinfo {note} {[Zh. Ehp. Teor. Fiz. 76, 1700
  (1979)]}\BibitemShut {NoStop}%
\bibitem [{\citenamefont {Wang}\ \emph {et~al.}(2019)\citenamefont {Wang},
  \citenamefont {Heinz}, \citenamefont {Verba}, \citenamefont {Kewenig},
  \citenamefont {Pirro}, \citenamefont {Schneider}, \citenamefont {Meyer},
  \citenamefont {L{\"a}gel}, \citenamefont {Dubs}, \citenamefont
  {Br{\"a}cher},\ and\ \citenamefont {Chumak}}]{Wang_PRL2019}%
  \BibitemOpen
  \bibfield  {author} {\bibinfo {author} {\bibfnamefont {Q.}~\bibnamefont
  {Wang}}, \bibinfo {author} {\bibfnamefont {B.}~\bibnamefont {Heinz}},
  \bibinfo {author} {\bibfnamefont {R.}~\bibnamefont {Verba}}, \bibinfo
  {author} {\bibfnamefont {M.}~\bibnamefont {Kewenig}}, \bibinfo {author}
  {\bibfnamefont {P.}~\bibnamefont {Pirro}}, \bibinfo {author} {\bibfnamefont
  {M.}~\bibnamefont {Schneider}}, \bibinfo {author} {\bibfnamefont
  {T.}~\bibnamefont {Meyer}}, \bibinfo {author} {\bibfnamefont
  {B.}~\bibnamefont {L{\"a}gel}}, \bibinfo {author} {\bibfnamefont
  {C.}~\bibnamefont {Dubs}}, \bibinfo {author} {\bibfnamefont {T.}~\bibnamefont
  {Br{\"a}cher}}, \ and\ \bibinfo {author} {\bibfnamefont {A.~V.}\ \bibnamefont
  {Chumak}},\ }\href {\doibase 10.1103/PhysRevLett.122.247202} {\bibfield
  {journal} {\bibinfo  {journal} {Phys. Rev. Lett.}\ }\textbf {\bibinfo
  {volume} {122}},\ \bibinfo {pages} {247202} (\bibinfo {year}
  {2019})}\BibitemShut {NoStop}%
\bibitem [{\citenamefont {Galkin}\ \emph {et~al.}(2006)\citenamefont {Galkin},
  \citenamefont {Ivanov},\ and\ \citenamefont {Zaspel}}]{Galkin_PRB2006}%
  \BibitemOpen
  \bibfield  {author} {\bibinfo {author} {\bibfnamefont {A.~Y.}\ \bibnamefont
  {Galkin}}, \bibinfo {author} {\bibfnamefont {B.~A.}\ \bibnamefont {Ivanov}},
  \ and\ \bibinfo {author} {\bibfnamefont {C.~E.}\ \bibnamefont {Zaspel}},\
  }\href {\doibase 10.1103/PhysRevB.74.144419} {\bibfield  {journal} {\bibinfo
  {journal} {Phys. Rev. B}\ }\textbf {\bibinfo {volume} {74}},\ \bibinfo
  {pages} {144419} (\bibinfo {year} {2006})}\BibitemShut {NoStop}%
\bibitem [{\citenamefont {Snyder}(1987)}]{Snyder_1987}%
  \BibitemOpen
  \bibfield  {author} {\bibinfo {author} {\bibfnamefont {J.~P.}\ \bibnamefont
  {Snyder}},\ }\href {\doibase 10.3133/pp1395} {\emph {\bibinfo {title} {{Map
  Projections: A Working Manual (U.S. Geological Survey Professional Paper
  1395)}}}},\ \bibinfo {type} {Tech. Rep.}\ (\bibinfo {year}
  {1987})\BibitemShut {NoStop}%
\bibitem [{\citenamefont {Naletov}\ \emph {et~al.}(2011)\citenamefont
  {Naletov}, \citenamefont {de~Loubens}, \citenamefont {Albuquerque},
  \citenamefont {Borlenghi}, \citenamefont {Cros}, \citenamefont {Faini},
  \citenamefont {Grollier}, \citenamefont {Hurdequint}, \citenamefont
  {Locatelli}, \citenamefont {Pigeau}, \citenamefont {Slavin}, \citenamefont
  {Tiberkevich}, \citenamefont {Ulysse}, \citenamefont {Valet},\ and\
  \citenamefont {Klein}}]{Naletov_PRB2011}%
  \BibitemOpen
  \bibfield  {author} {\bibinfo {author} {\bibfnamefont {V.~V.}\ \bibnamefont
  {Naletov}}, \bibinfo {author} {\bibfnamefont {G.}~\bibnamefont {de~Loubens}},
  \bibinfo {author} {\bibfnamefont {G.}~\bibnamefont {Albuquerque}}, \bibinfo
  {author} {\bibfnamefont {S.}~\bibnamefont {Borlenghi}}, \bibinfo {author}
  {\bibfnamefont {V.}~\bibnamefont {Cros}}, \bibinfo {author} {\bibfnamefont
  {G.}~\bibnamefont {Faini}}, \bibinfo {author} {\bibfnamefont
  {J.}~\bibnamefont {Grollier}}, \bibinfo {author} {\bibfnamefont
  {H.}~\bibnamefont {Hurdequint}}, \bibinfo {author} {\bibfnamefont
  {N.}~\bibnamefont {Locatelli}}, \bibinfo {author} {\bibfnamefont
  {B.}~\bibnamefont {Pigeau}}, \bibinfo {author} {\bibfnamefont {A.~N.}\
  \bibnamefont {Slavin}}, \bibinfo {author} {\bibfnamefont {V.~S.}\
  \bibnamefont {Tiberkevich}}, \bibinfo {author} {\bibfnamefont
  {C.}~\bibnamefont {Ulysse}}, \bibinfo {author} {\bibfnamefont
  {T.}~\bibnamefont {Valet}}, \ and\ \bibinfo {author} {\bibfnamefont
  {O.}~\bibnamefont {Klein}},\ }\href {\doibase 10.1103/PhysRevB.84.224423}
  {\bibfield  {journal} {\bibinfo  {journal} {Phys. Rev. B}\ }\textbf {\bibinfo
  {volume} {84}},\ \bibinfo {pages} {224423} (\bibinfo {year}
  {2011})}\BibitemShut {NoStop}%
\bibitem [{\citenamefont {Usov}\ and\ \citenamefont
  {Peschany}(1993)}]{Usov_JMMM1993}%
  \BibitemOpen
  \bibfield  {author} {\bibinfo {author} {\bibfnamefont {N.}~\bibnamefont
  {Usov}}\ and\ \bibinfo {author} {\bibfnamefont {S.}~\bibnamefont
  {Peschany}},\ }\href {\doibase 10.1016/0304-8853(93)90428-5} {\bibfield
  {journal} {\bibinfo  {journal} {J. Magn. Magn. Mater.}\ }\textbf {\bibinfo
  {volume} {118}},\ \bibinfo {pages} {L290} (\bibinfo {year}
  {1993})}\BibitemShut {NoStop}%
\bibitem [{\citenamefont {Guslienko}(2008)}]{Guslienko_JNN2008}%
  \BibitemOpen
  \bibfield  {author} {\bibinfo {author} {\bibfnamefont {K.~Y.}\ \bibnamefont
  {Guslienko}},\ }\href {\doibase 10.1166/jnn.2008.003} {\bibfield  {journal}
  {\bibinfo  {journal} {J. Nanosci. Nanotechnol.}\ }\textbf {\bibinfo {volume}
  {8}},\ \bibinfo {pages} {2745} (\bibinfo {year} {2008})}\BibitemShut
  {NoStop}%
\bibitem [{\citenamefont {Guslienko}\ \emph {et~al.}(2008)\citenamefont
  {Guslienko}, \citenamefont {Slavin}, \citenamefont {Tiberkevich},\ and\
  \citenamefont {Kim}}]{Guslienko_PRL2008}%
  \BibitemOpen
  \bibfield  {author} {\bibinfo {author} {\bibfnamefont {K.~Y.}\ \bibnamefont
  {Guslienko}}, \bibinfo {author} {\bibfnamefont {A.~N.}\ \bibnamefont
  {Slavin}}, \bibinfo {author} {\bibfnamefont {V.}~\bibnamefont {Tiberkevich}},
  \ and\ \bibinfo {author} {\bibfnamefont {S.-K.}\ \bibnamefont {Kim}},\ }\href
  {\doibase 10.1103/PhysRevLett.101.247203} {\bibfield  {journal} {\bibinfo
  {journal} {Phys. Rev. Lett.}\ }\textbf {\bibinfo {volume} {101}},\ \bibinfo
  {pages} {247203} (\bibinfo {year} {2008})}\BibitemShut {NoStop}%
\bibitem [{\citenamefont {Guslienko}\ and\ \citenamefont
  {Slavin}(2000)}]{Guslienko_JAP2000}%
  \BibitemOpen
  \bibfield  {author} {\bibinfo {author} {\bibfnamefont {K.~Y.}\ \bibnamefont
  {Guslienko}}\ and\ \bibinfo {author} {\bibfnamefont {A.~N.}\ \bibnamefont
  {Slavin}},\ }\href {\doibase 10.1063/1.372698} {\bibfield  {journal}
  {\bibinfo  {journal} {J. Appl. Phys.}\ }\textbf {\bibinfo {volume} {87}},\
  \bibinfo {pages} {6337} (\bibinfo {year} {2000})}\BibitemShut {NoStop}%
\bibitem [{\citenamefont {Ivanov}\ and\ \citenamefont
  {Wysin}(2002)}]{Ivanov_PRB2002}%
  \BibitemOpen
  \bibfield  {author} {\bibinfo {author} {\bibfnamefont {B.~A.}\ \bibnamefont
  {Ivanov}}\ and\ \bibinfo {author} {\bibfnamefont {G.~M.}\ \bibnamefont
  {Wysin}},\ }\href {\doibase 10.1103/PhysRevB.65.134434} {\bibfield  {journal}
  {\bibinfo  {journal} {Phys. Rev. B}\ }\textbf {\bibinfo {volume} {65}},\
  \bibinfo {pages} {134434} (\bibinfo {year} {2002})}\BibitemShut {NoStop}%
\bibitem [{\citenamefont {Taurel}\ \emph {et~al.}(2016)\citenamefont {Taurel},
  \citenamefont {Valet}, \citenamefont {Naletov}, \citenamefont {Vukadinovic},
  \citenamefont {de~Loubens},\ and\ \citenamefont {Klein}}]{Taurel_PRB2016}%
  \BibitemOpen
  \bibfield  {author} {\bibinfo {author} {\bibfnamefont {B.}~\bibnamefont
  {Taurel}}, \bibinfo {author} {\bibfnamefont {T.}~\bibnamefont {Valet}},
  \bibinfo {author} {\bibfnamefont {V.~V.}\ \bibnamefont {Naletov}}, \bibinfo
  {author} {\bibfnamefont {N.}~\bibnamefont {Vukadinovic}}, \bibinfo {author}
  {\bibfnamefont {G.}~\bibnamefont {de~Loubens}}, \ and\ \bibinfo {author}
  {\bibfnamefont {O.}~\bibnamefont {Klein}},\ }\href {\doibase
  10.1103/PhysRevB.93.184427} {\bibfield  {journal} {\bibinfo  {journal} {Phys.
  Rev. B}\ }\textbf {\bibinfo {volume} {93}},\ \bibinfo {pages} {184427}
  (\bibinfo {year} {2016})}\BibitemShut {NoStop}%
\bibitem [{\citenamefont {Ivanov}\ and\ \citenamefont
  {Zaspel}(2002)}]{Ivanov_APL2002}%
  \BibitemOpen
  \bibfield  {author} {\bibinfo {author} {\bibfnamefont {B.~A.}\ \bibnamefont
  {Ivanov}}\ and\ \bibinfo {author} {\bibfnamefont {C.~E.}\ \bibnamefont
  {Zaspel}},\ }\href {\doibase 10.1063/1.1499515} {\bibfield  {journal}
  {\bibinfo  {journal} {Appl. Phys. Lett.}\ }\textbf {\bibinfo {volume} {81}},\
  \bibinfo {pages} {1261} (\bibinfo {year} {2002})}\BibitemShut {NoStop}%
\bibitem [{\citenamefont {Zivieri}\ and\ \citenamefont
  {Nizzoli}(2005)}]{Zivieri_PRB2005}%
  \BibitemOpen
  \bibfield  {author} {\bibinfo {author} {\bibfnamefont {R.}~\bibnamefont
  {Zivieri}}\ and\ \bibinfo {author} {\bibfnamefont {F.}~\bibnamefont
  {Nizzoli}},\ }\href {\doibase 10.1103/PhysRevB.71.014411} {\bibfield
  {journal} {\bibinfo  {journal} {Phys. Rev. B}\ }\textbf {\bibinfo {volume}
  {71}},\ \bibinfo {pages} {014411} (\bibinfo {year} {2005})}\BibitemShut
  {NoStop}%
\bibitem [{\citenamefont {Buess}\ \emph {et~al.}(2005)\citenamefont {Buess},
  \citenamefont {Knowles}, \citenamefont {H{\"o}llinger}, \citenamefont {Haug},
  \citenamefont {Krey}, \citenamefont {Weiss}, \citenamefont {Pescia},
  \citenamefont {Scheinfein},\ and\ \citenamefont {Back}}]{Buess_PRB2005}%
  \BibitemOpen
  \bibfield  {author} {\bibinfo {author} {\bibfnamefont {M.}~\bibnamefont
  {Buess}}, \bibinfo {author} {\bibfnamefont {T.~P.~J.}\ \bibnamefont
  {Knowles}}, \bibinfo {author} {\bibfnamefont {R.}~\bibnamefont
  {H{\"o}llinger}}, \bibinfo {author} {\bibfnamefont {T.}~\bibnamefont {Haug}},
  \bibinfo {author} {\bibfnamefont {U.}~\bibnamefont {Krey}}, \bibinfo {author}
  {\bibfnamefont {D.}~\bibnamefont {Weiss}}, \bibinfo {author} {\bibfnamefont
  {D.}~\bibnamefont {Pescia}}, \bibinfo {author} {\bibfnamefont {M.~R.}\
  \bibnamefont {Scheinfein}}, \ and\ \bibinfo {author} {\bibfnamefont {C.~H.}\
  \bibnamefont {Back}},\ }\href {\doibase 10.1103/PhysRevB.71.104415}
  {\bibfield  {journal} {\bibinfo  {journal} {Phys. Rev. B}\ }\textbf {\bibinfo
  {volume} {71}},\ \bibinfo {pages} {104415} (\bibinfo {year}
  {2005})}\BibitemShut {NoStop}%
\bibitem [{\citenamefont {Vansteenkiste}\ \emph {et~al.}(2014)\citenamefont
  {Vansteenkiste}, \citenamefont {Leliaert}, \citenamefont {Dvornik},
  \citenamefont {Helsen}, \citenamefont {Garcia-Sanchez},\ and\ \citenamefont
  {{Van Waeyenberge}}}]{Vansteenkiste_AIPAdv2014}%
  \BibitemOpen
  \bibfield  {author} {\bibinfo {author} {\bibfnamefont {A.}~\bibnamefont
  {Vansteenkiste}}, \bibinfo {author} {\bibfnamefont {J.}~\bibnamefont
  {Leliaert}}, \bibinfo {author} {\bibfnamefont {M.}~\bibnamefont {Dvornik}},
  \bibinfo {author} {\bibfnamefont {M.}~\bibnamefont {Helsen}}, \bibinfo
  {author} {\bibfnamefont {F.}~\bibnamefont {Garcia-Sanchez}}, \ and\ \bibinfo
  {author} {\bibfnamefont {B.}~\bibnamefont {{Van Waeyenberge}}},\ }\href
  {\doibase 10.1063/1.4899186} {\bibfield  {journal} {\bibinfo  {journal} {AIP
  Advances}\ }\textbf {\bibinfo {volume} {4}},\ \bibinfo {eid} {107133}
  (\bibinfo {year} {2014}),\ 10.1063/1.4899186}\BibitemShut {NoStop}%
\bibitem [{\citenamefont {Verba}\ \emph {et~al.}(2018)\citenamefont {Verba},
  \citenamefont {Tiberkevich},\ and\ \citenamefont {Slavin}}]{Verba_PRB2018}%
  \BibitemOpen
  \bibfield  {author} {\bibinfo {author} {\bibfnamefont {R.}~\bibnamefont
  {Verba}}, \bibinfo {author} {\bibfnamefont {V.}~\bibnamefont {Tiberkevich}},
  \ and\ \bibinfo {author} {\bibfnamefont {A.}~\bibnamefont {Slavin}},\ }\href
  {\doibase 10.1103/PhysRevB.98.104408} {\bibfield  {journal} {\bibinfo
  {journal} {Phys. Rev. B}\ }\textbf {\bibinfo {volume} {98}},\ \bibinfo
  {pages} {104408} (\bibinfo {year} {2018})}\BibitemShut {NoStop}%
\end{thebibliography}

%

\end{document}